\documentclass[aps,superscriptaddress,floats,twocolumn,epsf,prr]{revtex4-2} 
\usepackage{graphicx}
\usepackage{epstopdf}
\usepackage[colorlinks=true,urlcolor=blue,citecolor=blue,linkcolor=blue,breaklinks=true]{hyperref}
\usepackage{color}
\usepackage{colortbl}
\usepackage{setspace}
\usepackage[utf8]{inputenc}
\usepackage{amsmath}
\usepackage{amssymb}
\usepackage{placeins}
\usepackage{mathtools}
\usepackage{tikz-feynman}
\usepackage{comment}
\usepackage[export]{adjustbox}

\usepackage[normalem]{ulem}

\usepackage{hyperref}
\hypersetup{colorlinks=true,breaklinks,linkcolor=blue,urlcolor=blue,citecolor=blue}


\begin{document}

\title{Disentangling real space fluctuations: the diagnostics of metal-insulator transitions beyond single-particle spectral functions}

\author{Michael Meixner}
\email{m.meixner@fkf.mpg.de}
\affiliation{Max-Planck-Institut f{\"u}r Festk{\"o}rperforschung, Heisenbergstra{\ss}e 1, 70569 Stuttgart, Germany}

\author{Marcel Kr{\"a}mer}
\affiliation{Max-Planck-Institut f{\"u}r Festk{\"o}rperforschung, Heisenbergstra{\ss}e 1, 70569 Stuttgart, Germany}
\affiliation{Institute of Information Systems Engineering, TU Wien, 1040 Vienna, Austria}
\affiliation{Institute of Solid State Physics, TU Wien, 1040 Vienna, Austria}

\author{Nils Wentzell}
\affiliation{Center for Computational Quantum Physics, Flatiron Institute, 162 5$^\text{th}$ Avenue, New York, NY 10010, USA}

\author{Pietro M. Bonetti}
\affiliation{Max-Planck-Institut f{\"u}r Festk{\"o}rperforschung, Heisenbergstra{\ss}e 1, 70569 Stuttgart, Germany}
\affiliation{Department of Physics, Harvard University, Cambridge MA 02138, USA}

\author{\\Sabine Andergassen}
\affiliation{Institute of Information Systems Engineering, TU Wien, 1040 Vienna, Austria}
\affiliation{Institute of Solid State Physics, TU Wien, 1040 Vienna, Austria}

\author{Alessandro Toschi}
\affiliation{Institute of Solid State Physics, TU Wien, 1040 Vienna, Austria}

\author{Thomas Sch{\"a}fer}
\email{t.schaefer@fkf.mpg.de}
\affiliation{Max-Planck-Institut f{\"u}r Festk{\"o}rperforschung, Heisenbergstra{\ss}e 1, 70569 Stuttgart, Germany}

\renewcommand{\i}{{\mathrm{i}}}
\date{ \today }

\begin{abstract} 
The destruction of metallicity due to the mutual Coulomb interaction of quasiparticles gives rise to fascinating phenomena of solid state physics such as the Mott metal-insulator transition and the pseudogap. A key observable characterizing their occurrences is the single-particle spectral function, determined by the fermionic self-energy. In this paper we investigate in detail how real space fluctuations are responsible for a self-energy that drives the Mott-Hubbard metal-insulator transition. To this aim we first introduce a real space fluctuation diagnostics approach to the Hedin equation, which connects the fermion-boson coupling vertex $\lambda$ to the self-energy $\Sigma$. Second, by using cellular dynamical mean-field theory calculations we unambiguously identify nearest-neighbor antiferromagnetic excitations as the leading physical processes responsible for the destruction of metallicity across the transition.
\end{abstract}
\maketitle

\section{Introduction}
\label{sec:introduction}
Strong correlations stemming from the mutual Coulomb interaction of electrons in a solid state system may not only lead to intriguing phenomena such as unconventional superconductivity \cite{Keimer2015,Scalapino2012}, quantum criticality \cite{Sachdev1999,Coleman2005}, and  metal-insulator transitions \cite{Imada1998}. They also pose great challenges to contemporary condensed matter theory since the description of strongly correlated systems cannot rely on an effective single-particle picture (for instance the density functional theory  \cite{Jones1989a}), which may only be successfully applied when such correlations are sufficiently weak. Hence, quantum many-body methods have to be employed for an in-depth understanding of these systems.

The arguably most prototypical example where the breakdown of a single-particle description occurs is the Mott-Hubbard metal-insulator transition (MIT) \cite{Mott1949,Mott1968}: there the thermodynamic transition between a metallic and an insulating state is driven by the strength of the Coulomb interaction, and, hence, the intensity of the correlations between electrically charged particles. This MIT has been observed experimentally in numerous materials, ranging, from transition-metal oxides \cite{McWhan1969,McWhan1970}, organic charge transfer salts \cite{Kanoda2011,Powell2006,Riedl2022}, to moir{\'e} transition-metal dichalchogenide heterostructures \cite{Li2021,Tscheppe2024} and two-dimensional metal-organic frameworks \cite{Lowe2024}. Further interest in Mott physics is triggered by the belief that the intriguing pseudogap phase, often observed in close vicinity to unconventional superconductivity \cite{Keimer2015,Riedl2022,Menke2024}, can be the result of doping a Mott insulator \cite{Lee2006}.

The simplest theoretical modelization of the MIT is realized in the single-band Hubbard model \cite{Qin2022,Arovas2022}, in which the Coulomb interaction is modelled as a purely local energy penalty $U$, which has to be paid whenever two electrons occupy the same lattice site. Such doubly occupied states are strongly suppressed in case of large values of the interaction strength, resulting in the formation of a gap  of width $U$ in the single-particle spectral function $A(\omega)$.

The first consistent description of the MIT in the Hubbard model has been one of the big early achievements of the dynamical mean-field theory (DMFT) \cite{Metzner1989,Georges1992a,Georges1996}. The MIT is described in DMFT as a first order transition with a second order critical endpoint. Hallmarks of this transition are (i) a vanishing quasiparticle weight $Z$, and (ii) the separation of two Hubbard subbands by width $U$ in the single-particle spectral function $A(\omega)$. As a mean-field theory in space, DMFT takes all local, temporal fluctuations into account, but neglects spatial correlations, crucially important for the description of the pseudogap physics. Calculations of the single-particle self-energy $\Sigma$, and, subsequently, of the Green function $G$ within diagrammatic \cite{Rohringer2018} and cluster \cite{Maier2005,Tremblay2006} extensions of DMFT can tell \textit{to which extent} the single-particle spectrum is modified by non-local correlations (see for instance \cite{Schaefer2015b,Schaefer2015c,Klett2020,Schaefer2021,Meixner2024,Sordi2012} for the half-filled Hubbard model on a square lattice).

For answering the question \textit{why} the spectral function exhibits certain properties [like being (pseudo-)gapped], and which physical processes are responsible for a specific single-particle response, the fluctuation diagnostics is the method of choice \cite{Gunnarsson2015,Schaefer2020}. The common thread of these methods is to access higher-order correlation functions out of which the contributions of effective scattering channels and their momenta and frequencies to the self-energy (in particular to the single-particle spectrum) of the system can be determined. Hitherto these methods have been successfully applied in momentum space for the pseudogap and superconducting regime of the two-dimensional Hubbard model, see for instance \cite{Gunnarsson2015,Gunnarsson2016,Wu2016,Schaefer2020,Arzhang2020,Dong2022}. The description, analysis, and impact of fluctuations, however, has rarely been investigated systematically from a real space perspective \cite{Gunnarsson2018} or for the MIT.

In this paper, we introduce a fluctuation diagnostic approach in real-space using the Hedin equation, which links the fermion-boson coupling vertex $\lambda$ to the fermionic self-energy $\Sigma$. We then calculate the components of the Hedin equation, i.e.~the Green function $G$, the bosonic propagator $w$, and $\lambda$ for the single-band Hubbard model on a square lattice by cellular dynamical mean-field theory (CDMFT) across the MIT on a cluster with $N_c\!=\!2\times 2$ sites. The center piece of this equation, the Hedin vertex $\lambda$, requires more computational effort for its calculation than single-particle quantities like $G$, and, therefore, has not been calculated before in real space CDMFT, but only momentum space DCA \cite{Kiese2024}. The Hedin vertex contains significantly more information, namely how the system's fermionic constituents couple to bosonic (charge and spin) modes in space and in time. The application of the real space fluctuation diagnostics, as we will show, allows (i) to trace back the origin of the lower critical interaction at which the MIT occurs, w.r.t. (single-site) DMFT, (ii) to determine the role of the (non-local) fermion-boson coupling vertex, and (iii) to pin-point the diagrams (and, hence, physical processes) in real space which drive the opening of the MIT gap.
 
The paper is arranged as follows: Sec.~\ref{sec:Model} introduces the Hubbard model and CDMFT. In Sec.~\ref{sec:fluctuationdiagnostics} we derive the real space fluctuation diagnostics procedure using the Hedin equation with the fermion-boson vertex. In Sec.~\ref{sec:single-particle} we recapitulate the single-particle charateristics of the MIT. Sec.~\ref{sec:sig_loc}, in a first step, analyzes the MIT by focusing on the on-site self-energy and then disentangle the real-space contributions by connecting the self-energy to diagrammatic real space spin- and charge contributions between specific cluster sites. In a second step we relate these quantitative results to qualitative features of the fermion-boson coupling vertex. Subsequently, a similar analysis is conducted for the non-local cluster self-energies in Sec.~\ref{sec:sig_nonloc}. The results are then summarized, connected to momentum-space quantities and set into context in Sec.~\ref{sec:conclusion}. The raw data of the presented figures is publicly available at \cite{DataEdmond}.

\section{The Hubbard model and cellular dynamical mean-field theory} \label{sec:Method}
\label{sec:Model}
We study the particle-hole symmetric, two-dimensional Hubbard model \cite{Hubbard1963,Hubbard1964,Gutzwiller1963,Kanamori1963,Qin2022,Arovas2022} on the simple (unfrustrated) square lattice 
\begin{equation} \label{eq:Hamiltonian}
H=-t\sum_{\langle i,j \rangle,\sigma}c^\dagger_{i,\sigma}c_{j,\sigma}+U\sum_{i}n_{i,\uparrow}n_{i,\downarrow}-\frac{U}{2} \sum_{i,\sigma}n_{i,\sigma},
\end{equation}
where $c_{i,\sigma}^\dagger$ ($c_{i,\sigma}$) represent the site-dependent fermionic creation (annihilation) operators of an electron with spin $\sigma\in\{\uparrow,\downarrow\}$ on site $i$, $t$ is the nearest-neighbor (n) tunneling amplitude (hopping) and $U$ the local Coulomb repulsion. 
\subsection{Cellular dynamical mean-field theory}
We solve the Hubbard model in the approximation of the cellular dynamical mean-field theory (CDMFT) \cite{Kotliar2001,Maier2005}. CDMFT, in comparison to DMFT, includes short range fluctuations up to the cluster size $N_c$ exactly, and is a controlled approximation in this quantity, i.e., it yields the exact solution for $N_c\!\rightarrow\!\infty$. In this work we focus on an impurity with $N_c\!=\!2\times 2$ sites. In CDMFT this cluster is embedded in a fermionic bath, i.e., the original model is mapped onto an effective Anderson impurity problem, which is solved self-consistently. Details of the algorithm are given in App.~\ref{App.CDMFT-Algorithm} \cite{Meixner2024}. We stress here that the fermionic bath's coupling to the impurity is only energy dependent, i.e., the approximation allows for fermionic fluctuations between the impurity and the bath, while real space correlations stemming from the mutual Coulomb interaction are taken into account only within the impurity. 
The auxiliary Anderson model is solved via the interaction expansion continuous-time quantum Monte-Carlo (CT-INT) solver provided as part of the TRIQS library \cite{TRIQS}. For illustrative purposes, spectral functions provided are obtained from the maximum entropy analytic continuation algorithm (MaxEnt) \cite{Silver1990,Bryan1990,Kraberger2017} delivered by TRIQS \footnote{\url{https://triqs.github.io/maxent/latest/guide/tau_maxent.html}; See: \cite{TRIQS}}.

For comparison, we show calculations for the Hubbard atomic cluster (i.e., a cluster without an embedding), which were obtained from a Lehmann representation for two- and three-point functions \cite{Tagliavini2018}.

\section{Real space fluctuation diagnostics for the Hubbard model}
\label{sec:fluctuationdiagnostics}
The self-energy of a system can be obtained via one of the Hedin equations \cite{Hedin1965,Biermann2008,Krien2019,Patricolo2024}. In the Matsubara formalism, and for the real-space self-energy $\Sigma^{s,e}(\nu_n)$ between two cluster sites $s$ and $e$, we find:
\begin{equation}
\label{eq:Hedin}
\begin{split}
    \Sigma&^{s,e}(\i\nu_n)=T\sum_{\omega_m}\tilde{\Sigma}^{s,e}_{\text{sp/ch}}(\i\omega_m,\i\nu_n)\\&\coloneqq \pm UT\sum_{\omega_m}\sum_{b,f}G^{s,f}\left(\i\omega_m+\i\nu_n\right)w^{b,s}_{\text{sp/ch}}\left(\i\omega_m\right)\lambda_{\text{sp/ch}}^{b,f,e}\left(\i\omega_m,\i\nu_n \right )\\&\quad+\delta^{s,e}\Sigma_{C,\mathrm{sp/ch}},
\end{split}
\end{equation}
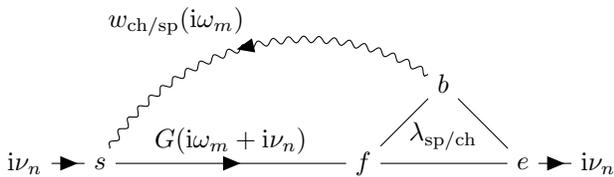
\begin{figure}[t!]
    \centering
\begin{tikzpicture} 
\begin{feynman}
\vertex (i) {$s$}; 
\vertex [below=0.4cm of i] (belowi); 
\vertex [right=3.5cm of i] (n) {$f$}; 
\vertex [above right=of n] (m) {$b$};
\vertex [below=0.7cm of m] (lambda) {$\lambda_{\text{sp/ch}}$};
\vertex [below right=of m] (j) {$e$};
\vertex [below=0.4cm of j] (belowj); 
\vertex (in) [left=1cm of i]{$\mathrm{i}\nu_n$};
\vertex (out) [right=1cm of j]{$\mathrm{i}\nu_n$};
\diagram* {(i) -- [fermion, edge label=\( G(\mathrm{i}\omega_m+\mathrm{i}\nu_n)\)] (n) -- (j) -- (m)--(n),
(i) --[anti charged boson, quarter left, edge label=\( w_{\text{ch/sp}}(\mathrm{i}\omega_m) \)](m),
(in) -- [fermion](i),
(j) -- [fermion](out),
}; 
\end{feynman}
\end{tikzpicture}
    \caption{Schematic representation of the real space Hedin equation (\ref{eq:Hedin}), where real space points $s,e,f,b$ are indicated by italic letters. To obtain the full impurity self-energy between start and end site $s,e$, the summation over the internal coordinates of the fermionic and bosonic end $f,b$ of the Hedin vertex $\lambda^{\mathrm{b,f,e}}$ has to be carried out. This plot has been created with \cite{Tikz}.}
    \label{fig:Hedin}
\end{figure}a Feynman diagrammatic illustration of which is given in Fig.~\ref{fig:Hedin}. Here $G$ represents the fermionic propagator. $w_{\text{sp/ch}}$ describes the renormalized propagation of charge or spin bosons, including the bare interaction $U$. The central quantity of this equation is given by $\lambda_{\text{ch/sp}}$, the fermion-boson Hedin vertex: it encapsulates scattering process between a fermion and a boson in the charge or spin channel, respectively. $\Sigma_{C,\mathrm{sp/ch}}$, is a real, constant, i.e.~frequency independent Hartree-like contribution for the respective channel [for details and derivation see App.~\ref{App:Def}, particularly Eq.~(\ref{eq:fullhedin})]. The formulation of Eq.~(\ref{eq:Hedin}) is equivalent in the physical channels, such that performing the sums on the right hand side of the equation yields the same self-energy. $\omega_m=2m\pi T$ are bosonic Matsubara frequencies,  $\nu_n=(2n+1)\pi T$ fermionic ones (with $m,n \in \mathbb{Z}$). Please note that $+$ represents the spin channel and $-$ the charge channel, respectively. Since in the particle-hole symmetric situation under study, where particle-hole fluctuations are expected to dominate the physics, we omit the corresponding analysis in the pairing channel formulation. The external real space coordinates of the self-energy are labelled $s,e$, representing the start and end points of a diagram. Internal orbitals are labelled $b,f$ to indicate that they connect the bosonic and fermionic end points of the Hedin vertex. We note that Eq.~(\ref{eq:Hedin}) is exact if all the constituents of this equation are known exactly. By setting $\lambda_{\text{sp/ch}}=1$ (corresponding to its high-frequency limit) one recovers the well-known GW approximation (see, e.g., \cite{Held2011}).

\subsection{Fluctuation diagnostics with the Hedin vertex\texorpdfstring{ $\lambda$}{}} 
We diagnose the internal bosonic fluctuations of the self-energy in the respective channel by analyzing $\tilde{\Sigma}^{s,e}_{\text{ch/sp}}(\i\omega,\i\nu)$, where the tilde indicates that the bosonic Matsubara sum over $\omega_m$ has not yet been executed in the Hedin Eq.~(\ref{eq:Hedin}). In this respect, our approach is similar to earlier works in fluctuation diagnostics \cite{Gunnarsson2015,Schaefer2020} formulated for the Schwinger-Dyson equation. There the basic building brick for the analysis is the four-point function (two-particle fermionic scattering amplitude), a much heavier object to compute due to its dependence on three frequencies/times and three momenta/site-indices. Here, instead, we extract the relevant information from the three-point function, i.e., the fermion-boson coupling vertex $\lambda$, which reduces memory and computational cost and allows for the consideration of real space clusters. By using $\lambda$, we can discriminate the contributions of static and dynamic fermion-boson scattering processes on the cluster impurity in the respective physical channel to the full self-energy.

\subsection{Real space fluctuation diagnostics} 
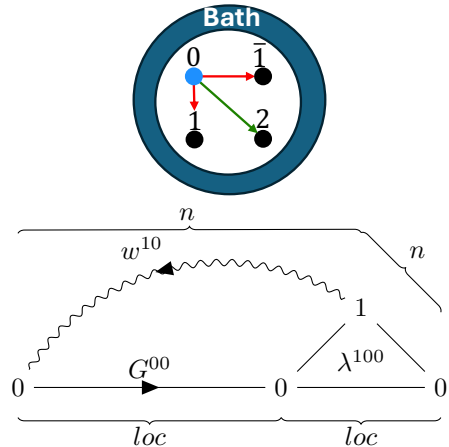
\begin{figure}[t!]
    \centering
\includegraphics[width=0.3\linewidth]{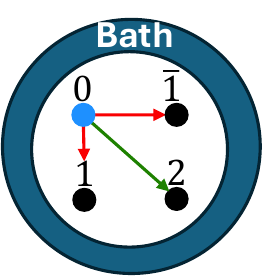}
\begin{tikzpicture} 
\begin{feynman}
\vertex (i) {$\mathrm{0}$}; 
\vertex [below=0.4cm of i] (belowi); 
\vertex [above=2.05cm of i] (abovei); 
\vertex [right=3.5cm of i] (n) {$\mathrm{0}$}; 
\vertex [above right=of n] (m) {$\mathrm{1}$};
\vertex [below=0.7cm of m] (lambda) {$\lambda^{\mathrm{100}}$};
\vertex [below right=of m] (j) {$\mathrm{0}$};
\vertex [below=0.4cm of j] (belowj); 
\vertex [above=1cm of j] (abovej);
\vertex [above=1cm of m] (aboveb);
\vertex [right=0.05cm of m] (aboveb2);
\vertex [below=0.4cm of n] (belowf); 
\diagram* {(i) -- [fermion, edge label=\(G^{\mathrm{00}}\)] (n) -- (j) -- (m)--(n),
(i) --[anti charged boson, quarter left, edge label=\( w^{\mathrm{10}}\)](m),
}; 
\draw [decoration={brace}, decorate] (belowj.south) -- (belowf.south) node [pos=0.5, below] {\(loc\)};
\draw [decoration={brace}, decorate] (belowf.south) -- (belowi.south) node [pos=0.5, below] {\(loc\)};
\draw [decoration={brace}, decorate] (aboveb.north) -- (abovej.north) node [pos=0.49, above right] {\(n_{\,}\)};
\draw [decoration={brace}, decorate] (abovei.north) -- (aboveb.north) node [pos=0.5, above] {\(n_{\,}\)};
\end{feynman}
\end{tikzpicture}
    \caption{Sketch of the labelling of the impurity sites and illustration of the real space analysis of the summand Eq.~(\ref{eq:Hedin_example}) of the Hedin equation for the local self-energy. As the self-energy is local, this implies the external start and end orbitals $s,e$ to be equal. Here we illustrate the summand, where the bosonic propagator covers a nearest-neighbor distance $w^\mathrm{10}$ and the fermionic propagator is local $G^\mathrm{00}$. The braces indicate the real space distances covered between start and end orbitals $s,e,b,f$, on the cluster which are, for this example, either local ($loc$) or nearest-neighbor ($n$). This plot has been created with \cite{Tikz}.}
    \label{fig:Hedin_example}
\end{figure}

The self-energy can be analyzed in even more detail by considering each summand of the internal real space sum over the orbitals $b$ and $f$ separately. To that end, we label the impurity sites $\mathrm{0,1,\overline{1},2}$, where we start at the top left hand side as is illustrated in the upper panel of Fig.~\ref{fig:Hedin_example}. This allows to disentangle the contributions from an on-site bosonic fluctuation and specific non-local bosonic fluctuations: For instance, if the bosonic propagator is local, i.e., the sites $b=s=\mathrm{0}$, we refer to this contribution as the ``bosonically local" ($w^{b,s}=w^{\mathrm{00}}$, blue in Fig.~\ref{fig:Hedin_example}) contribution to the self-energy, which contrasts the ``non-local" contributions, such as the ``bosonic nearest neighbor" contribution ($w^{\mathrm{10}},w^{\mathrm{\overline{1}0}}$, red in Fig.~\ref{fig:Hedin_example}) or ``bosonic second nearest neighbor" ($w^{\mathrm{20}}$, green in Fig.~\ref{fig:Hedin_example}) contribution. Note, that due to restriction to the paramagnetic case in this study, rotational and mirror symmetries of the cluster translate directly to involved quantities $G,w,\lambda$. Hence, these functions are equal for geometrically equivalent combinations of their leg endpoints. For example, for the bosonic propagator, we have eight next-neighbor combinations: $w^{\mathrm{01}}=w^{\mathrm{0\overline{1}}}=w^{\mathrm{12}}=w^{\mathrm{\overline{1}2}} + \mathrm{transposed}$. 

This real-space analysis can then be combined with the aforementioned fluctuation diagnostics, by analyzing the bosonic fluctuations which cover a certain distance within the cluster. We coin this approach the ``real-space fluctuation diagnostics" (RFD). An analogous classification can be applied for the fermionic propagator $G$. Fixing the distance covered by a bosonic and a fermionic propagator for a certain self-energy automatically leaves only one three-point Hedin vertex contributing to this specific summand.

Let us give an instructive example, which is illustrated in 
Fig.~\ref{fig:Hedin_example}: 
 If we aim at the on-site self-energy $\Sigma^{\mathrm{00}}$, and analyze the contribution of the local fermionic propagator   $G^{\mathrm{00}}$, this would fix the orbitals to $f=s=e=0$. Further, we want to inspect the nearest-neighbor bosonic fluctuations $w^{\mathrm{10}}$, i.e., the case of $b=1$. This results in a Hedin vertex $\lambda^{b,f,e}=\lambda^{\mathrm{100}}$ and the resulting self-energy component $\tilde{\Sigma}^{\mathrm{00}}\left[\lambda^{\mathrm{100}}\right]$, where the tilde indicates that the bosonic sum has not yet been carried out:
 \begin{equation}
\label{eq:Hedin_example}
\begin{split}
\tilde{\Sigma}^{\mathrm{00}}&\left[\lambda^{\mathrm{100}}\right]\left(\i\omega_m,\i\nu_n\right)\\&=\pm UG^{\mathrm{00}}\left(\i\omega_m+\i\nu_n\right)w^\mathrm{10}_{\text{sp/ch}}\left(\i\omega_m\right)\lambda_{\text{sp/ch}}^{\mathrm{100}}\left(\i\omega_m,\i\nu_n \right ).
\end{split}
\end{equation}
This Feynman diagram is shown in Fig.~\ref{fig:Hedin_example}.

In the following we analyze $\tilde{\Sigma}^{s,e}(\i\omega_m,\i\nu_n)$ and, in the spirit of fluctuation diagnostics, we refer to a sharp bosonic frequency structure of its diagrams as ``peaked", contrasted by the term ``broad" for a rather flat distribution.

\section{Single-Particle Characteristics of the MIT}
\label{sec:single-particle}

\begin{figure}[t!]
    \centering
    \includegraphics[width=\linewidth,right]{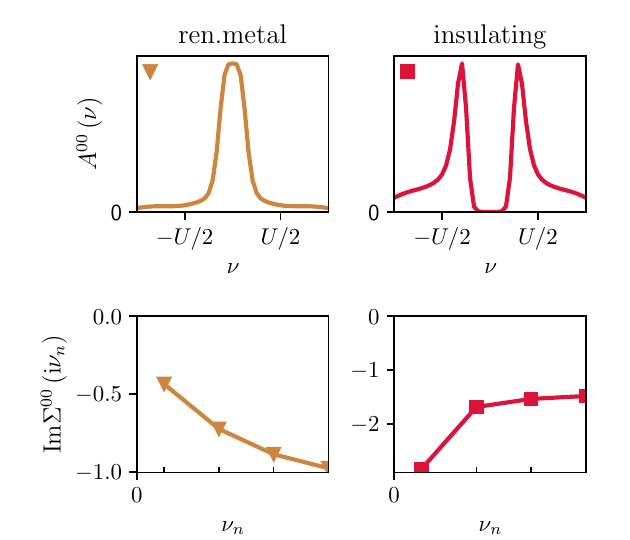}
    \includegraphics[width=0.99\linewidth]{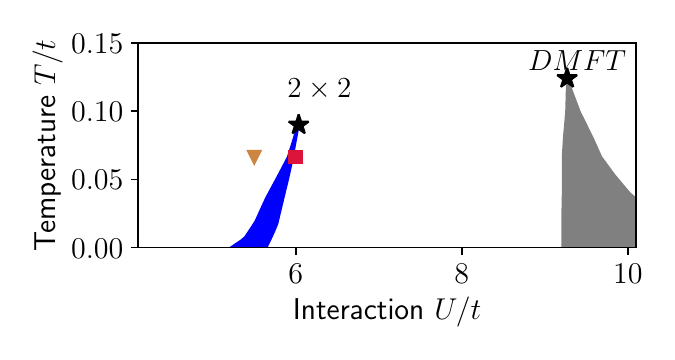}
    \caption{Upper row: local spectral functions computed with $2\times2$-CDMFT for the parameters marked in the lower row (brown triangle for the renormalized metallic and red square for the insulating spectral function, respectively): $U=5.5t$ (left) and $U=6t$ (right) at $T=0.067t$. Central row: corresponding imaginary parts of the on-site self-energies as a function of the Matsubara frequency, where inward marker give the respective Matsubara frequency and the zero-frequency argument is highlighted by $0$. Lower row: phase diagram of the half-filled DMFT \cite{Pelz2023} (grey) and $2\times2$-CDMFT \cite{Park2008} (blue) giving the first order MIT coexistency region. The second order critical endpoint is marked by a black star.}
    \label{fig:Phasediagram}
\end{figure}

Before we use the real space fluctuation diagnostics for the analysis of the Mott metal-insulator transition, we recapitulate the results from (single-site, $N_c\!=\!1$) DMFT and CDMFT on a $N_c\!=\!2\times 2$ cluster on the single-particle level.

One of the big early successes of DMFT has been its consistent description of an interaction-driven metal-insulator transition in the paramagnetic phase of the Hubbard model \cite{Georges1996,Kotliar2004,Georges1993,Bluemer2002}. 
At low values of the Hubbard interaction $U$ the purely local fluctuations considered by DMFT lead to a renormalization of the quasiparticle parameters in the metallic regime, while at larger values of the interaction, a gap in the single-particle spectral function and, thus, Mott-insulating regime arises. On the level of the self-energy, the gap is produced by a divergence (either on the real frequency or Matsubara axis) of the imaginary part of the impurity self-energy. The metallic and Mott-insulating regimes are separated by a first order MIT and an area of hysteresis, where the metallic and insulating phase coexist.
This coexistence region is shown for the Hubbard model on a simple square lattice in the (paramagnetically restricted) DMFT solution in grey in the phase diagram depicted in the lower row of Fig.~\ref{fig:Phasediagram}. Its critical endpoint (marked by a black star) is located at $U_{c}^{\mathrm{DMFT}}\approx 9.3t$ and $T_{c}^{\mathrm{DMFT}}\approx0.125t$. When enlarging the impurity to a $2\times2$-CDMFT cell to incorporate short-ranged fluctuations, the critical interaction is strongly reduced to $U_c^{\mathrm{2\times 2}}\approx 6t$ at $T_c^{\mathrm{2\times 2}}\approx 0.09/t$ (blue coexistence region with a black star), separating again a strongly renormalized metallic regime from an insulating regime \cite{Parcollet2004,Stanescu2006,Park2008,Fratino2017,Walsh2019,Walsh2019b}. The spectrum of a strongly renormalized metal, where the local spectral function \begin{equation}
A^{\mathrm{00}}(\nu)=-\frac{1}{2\pi}\left.\mathrm{Im }G^{\mathrm{00}}(\i\nu)\right|_{\i\nu\rightarrow\nu}
\end{equation} displays a clear quasi-particle peak, is shown in Fig.~\ref{fig:Phasediagram} (upper row) for $U=5.5t$ and $T=0.067t$, and, in contrast to that, a data point at slightly higher interaction $U=6t$, just beyond the metal-insulator transition. The spectral function of the insulating data point displays a clear gap with a width of order $U$. The corresponding imaginary parts of the on-site self-energies are presented in the central row of Fig.~\ref{fig:Phasediagram}. We further stress that the width of the coexistence region is reduced significantly by the inclusion of non-local correlations on top of DMFT. Let us comment that in this work we consider the $N_c\!=\!2 \times 2$ cluster only. For the dependence on cluster size and geometry see \cite{Sakai2012,Meixner2024,Downey2023}.

In the following, we first analyze the on-site self-energy in the strongly renormalized metallic regime and contrast it to the insulating phase in close proximity to the MIT (see brown triangle and red square in Fig.~\ref{fig:Phasediagram}, respectively) to disentangle the relevance of local and non-local bosonic and fermionic fluctuations in Sec.~\ref{sec:sig_loc}. Afterwards, we discuss the decomposition of the non-local self-energies and their effect on the fermionic spectral properties in Sec.~\ref{sec:sig_nonloc}.

\section{Decomposing the on-site self-energy\texorpdfstring{ $\Sigma_{\mathrm{sp/ch}}^\mathrm{00}$}{}}
\label{sec:sig_loc}
\begin{figure}[t!]
    \centering
    \includegraphics[width=\linewidth,right]{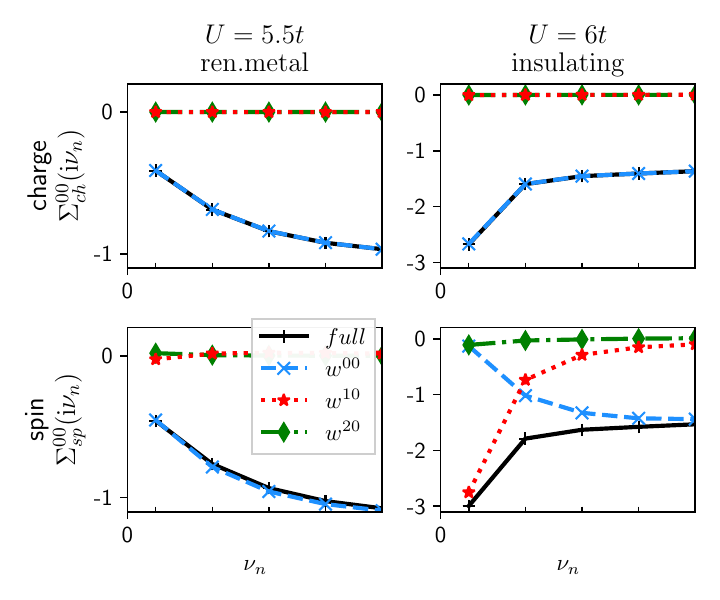}
    \caption{Self-energies computed via CDMFT for a strongly renormalized metal at $U=5.5t$ (left column) and an insulator at $U=6t$ and $T=0.067t$. Self-energies $\Sigma^\text{00}$ (black solid line) and all contributions to it from different sets of diagrams in the charge- and spin channel are given in the first and second row in the charge and spin picture, respectively. Here, we distinguish the contributions arising from the bosonic local $w^{\mathrm{00}}$ (blue dashed line), the bosonic nearest-neighbor $w^\mathrm{10}$ and $w^\mathrm{\overline{1}0}$ (red dotted line) and the bosonic second-nearest neighbor $w^{\mathrm{20}}$ diagrams (green dot-dashed line) contributing to the full self-energy.}
    \label{fig:local_analysis_Siglocal}
\end{figure}

\begin{figure*}[t!]
    \centering
    \begin{tikzpicture} 
\begin{feynman}
\vertex (i) {$\mathrm{0}$}; 
\vertex [left=4.15cm of i] (i0){ }; 
\vertex [right=.8cm of i] (n) {$f$}; 
\vertex [above right=0.8cm of n] (m) {$\mathrm{0}$};
\vertex [below right=0.8cm of m] (j) {$\mathrm{0}$};
\diagram* {(i) -- [fermion] (n) -- (j) -- (m)--(n),
(i) --[anti charged boson, quarter left](m),
}; 
\vertex [right=3.47cm of i] (i2) {$\mathrm{0}$}; 
\vertex [right=.8cm of i2] (n2) {$f$}; 
\vertex [above right=0.8cm of n2] (m2) {$\mathrm{1}$};
\vertex [below right=0.8cm of m2] (j2) {$\mathrm{0}$};
\diagram* {(i2) -- [fermion] (n2) -- (j2) -- (m2)--(n2),
(i2) --[anti charged boson, quarter left](m2),
}; 
\vertex [right=3.47cm of i2] (i3) {$\mathrm{0}$}; 
\vertex [right=.8cm of i3] (n3) {$f$}; 
\vertex [above right=0.8cm of n3] (m3) {$\mathrm{\overline{1}}$};
\vertex [below right=0.8cm of m3] (j3) {$\mathrm{0}$};
\diagram* {(i3) -- [fermion] (n3) -- (j3) -- (m3)--(n3),
(i3) --[anti charged boson, quarter left](m3),
}; 
\vertex [right=3.47cm of i3] (i4) {$\mathrm{0}$}; 
\vertex [right=.8cm of i4] (n4) {$f$}; 
\vertex [above right=0.8cm of n4] (m4) {$\mathrm{2}$};
\vertex [below right=0.8cm of m4] (j4) {$\mathrm{0}$};
\diagram* {(i4) -- [fermion] (n4) -- (j4) -- (m4)--(n4),
(i4) --[anti charged boson, quarter left](m4),
}; 
\end{feynman}
\end{tikzpicture}
    \includegraphics[width=1\linewidth]{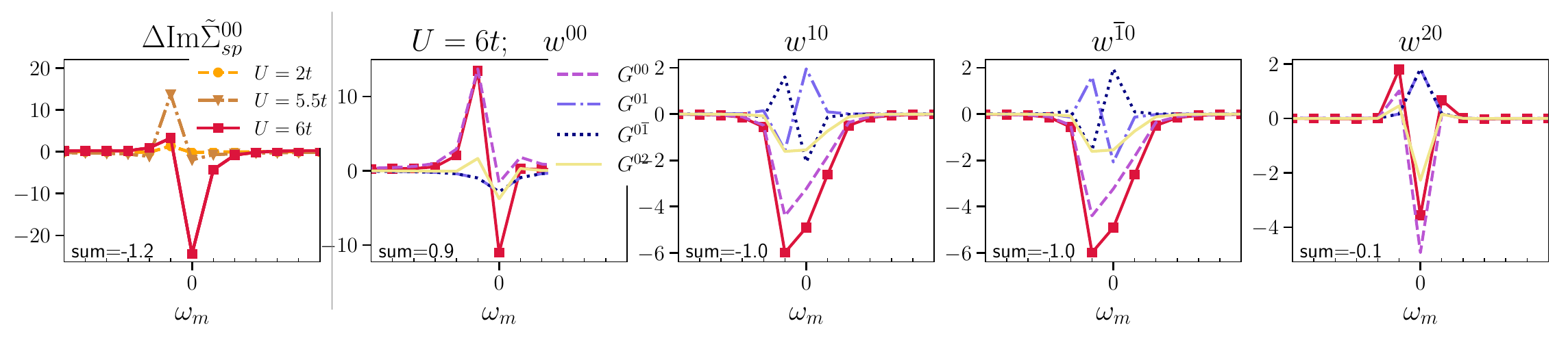}
    \caption{Real space fluctuation diagnostics of the imaginary part of the local self-energy's difference. The left panel shows the sum over all orbitals for different interactions. The four right panels show the respective real space contributions to the Hedin equation for a fixed bosonic propagator $w$ for $U=6t$ as a red solid line (full squares). For each bosonic orbital, all contributions from different fermionic propagators are given in the respective plot, beginning by the local fermionic propagator as the pink dashed line. ``sum" gives the overall sum over all data points for $U=6t$ for the respective case.}
    \label{fig:FD_sigma_local}
\end{figure*}

In single-site DMFT the quasi-particle peak in terms of the Matsubara formalism is encoded by the (imaginary part of) a self-energy which has a positive difference of first and second Matsubara frequency \cite{Arzhang2020}:
\begin{equation}
\Delta{\Sigma}\coloneqq\Sigma(\i\nu_{n=0})-\Sigma(\i\nu_{n=1}).
\end{equation}
It, thus, can be Taylor expanded around the frequency $\i\nu=0$. The spectral function in the insulating regime on the other hand is suppressed by a self-energy with a negative difference, indicating a pole of the on-site self-energy at $\i\nu=0$. In this section, we first inspect the on-site self-energy $\Sigma^\mathrm{00}$ calculated within CDMFT by means of a real space fluctuation diagnostics analysis. We approach the problem in three steps: we analyze (i) the contributions of different frequencies of the bosonic propagator to the self-energy (Fig.~\ref{fig:FD_Siglocal}), (ii) the impact of different bosonic distances to the full self-energy (Fig.~\ref{fig:local_analysis_Siglocal}), (iii) the full real space fluctuation diagnostics (Fig.~\ref{fig:FD_sigma_local}), and, (iv) the impact of the structure of the Hedin vertex (Fig.~\ref{fig:Hedin-to-sigma_local}) on the self-energy over the MIT.
\subsection{Bosonic frequency analysis of the on-site self-energy}
\begin{figure}[t!]
    \centering
    \includegraphics[width=\linewidth]{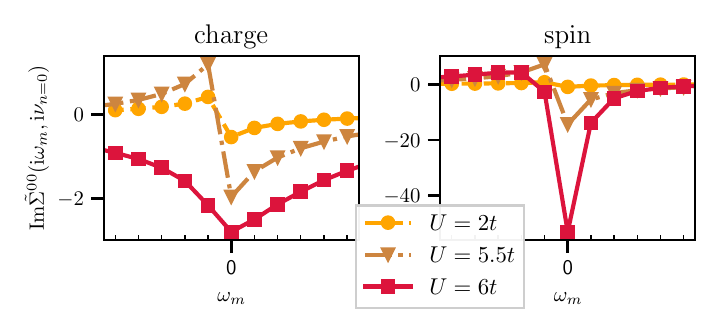}
    \caption{Bosonic contributions to the Hedin equation for the imaginary part of $\Sigma^{\mathrm{00}}(\i\nu_{n=0})$ at the zeroth fermionic Matsubara frequency in the charge (left panel) and spin channel (right panel), respectively, at $T=0.067t$. Please note the different scales. The self-energies at $U=5.5t$ and $U=6t$ correspond to the ones shown in the central row of Fig.~\ref{fig:Phasediagram}.}
    \label{fig:FD_Siglocal}
\end{figure}
We first analyze the differences between the metallic and insulating regimes by omitting the bosonic sum in the Hedin equation for the self-energies discussed earlier in Fig.~\ref{fig:Phasediagram}, however keeping the orbital sum in place:
\begin{equation}
\label{eq:Hedin_fluc}
\begin{split}
    \tilde{\Sigma}^{0,0}_{\text{sp/ch}}(\i\omega_m,\i\nu_{n=0})&\coloneqq\Sigma_C\pm \sum_{b,f}G^{0,f}\left(\i\omega_m+\i\nu_{n=0}\right)\\ &\cdot \lambda_{\text{sp/ch}}^{b,f,0}\left(\i\omega_m,\i\nu_{n=0} \right )w^{b,0}_{\text{sp/ch}}\left(\i\omega_m\right).
    \end{split}
\end{equation}
The horizontal axis of Fig.~\ref{fig:FD_Siglocal} gives the bosonic transfer frequency carried by the respective bosonic propagator in Eq.~(\ref{eq:Hedin_fluc}) for the different interaction values. Negative values contribute to the overall negative and insulating self-energy (`damping'), while positive values result in a more metallic self-energy (`anti-damping').
In the charge channel (left column), and for the metallic cases at $U\!=\!2t$ and $U\!=\!5.5t$ (orange circles and brown  triangles, respectively), we observe a change of sign between the minus-first and zeroth bosonic frequency, leading to a cancellation when executing the bosonic Matsubara sum, and hence a suppression of the self-energy. In the insulating regime ($U\!=\!6t$, red squares) we find a broad, overall negative structure. In contrast, in the spin channel (second column), only the lowest frequencies contribute, at least for high interaction values and, in particular, a prominent negative peak emerges for the lowest bosonic frequency in the insulating case $U=6t$. This indicates that a static (or, at most, low-frequency) spin mode is sufficient to describe the insulating nature of the system.

This is in accordance with a fluctuation analysis of the single-site DMFT, where the static spin contribution at $\i\omega=0$ displays a strong peak as a contribution to the self-energy, while the charge diagrams in the insulating regime, similar to CDMFT in Fig.~\ref{fig:FD_Siglocal}, displays a broad Lorentzian-like shape. For a detailed analysis of the DMFT case see App.~\ref{App:DMFT}. While in DMFT, only local diagrams contribute to the self-energy, we aim now at disentangling the influence of real-space fluctuations to the on-site self-energy in CDMFT.

\subsection{Real space analysis of the local self-energy: The bosonic contribution}

The Hedin equation (\ref{eq:Hedin}) allows for a decomposition of the corresponding self-energy, represented either in the spin or charge channel $\Sigma^\mathrm{00}_\text{sp/ch}$, into local and non-local contributions. We here classify contributions to the Hedin equation in the specific channel by differing between the distance covered by the bosonic propagator $w$ when executing the bosonic orbital sum over the orbital $b$ in Eq.~(\ref{eq:Hedin}). While the result of this sum in the Hedin equation amounts to the full self-energy (in both channels), the weight of local and non-local diagrams are not necessarily evenly distributed. This bosonic real space classification leads to three classes of diagrams: (i) the diagram where the bosonic propagator is local $w^\mathrm{00}$ (``bosonically local''), (ii) where the bosonic propagator covers the nearest-neighbor distance $w^\mathrm{10}$, and (iii) the second-nearest neighbor distance $w^\mathrm{20}$.

The contributions of these real space classes of diagrams to the local self-energy \footnote{While the expression of the self-energy via the Hedin equation in different channels is formally equivalent, slight deviations can be noticed. This is caused by a Matsubara summation over a finite frequency grid. However, the error made is not qualitative and only of small quantitative nature, see App.~\ref{App:Error}.} in the respective channel is displayed in Fig.~\ref{fig:local_analysis_Siglocal}. In the charge channel (central row), the self-energy (black solid line) is almost solely constituted by the bosonically local ($w^\mathrm{00}_{\text{ch}}$, blue dashed line) diagrams in both, the metallic and insulating phase. The strong pole-like increase of the self-energy in the insulating phase originates from the bosonically local charge diagrams. This contrasts with the spin Hedin equation (second row): There, the self-energy in the metallic case is constituted exclusively by the bosonically local diagrams (dashed blue line). However, conversely to the charge channel, this class of diagrams does not gain in strength when turning into the insulating regime, but rather \textit{reduces} its contribution to the self-energy at the lowest Matsubara frequency when increasing the interaction. This reduction of the self-energy is overcompensated by the bosonic nearest-neighbor ($w^\mathrm{10}_{\text{sp}}$, red dotted line) diagram, which, initially hardly contributing for $U=5.5t$, does now contribute strongly for the lowest Matsubara frequency and generates the insulating pole of the full self-energy. The second-nearest neighbor diagram ($w^\mathrm{20}_{\text{sp}}$, dotted-dashed line) still remains negligible.

The locality of the charge diagrams can be straightforwardly explained by the site dependence of the bosonic charge propagator: For $w_{\mathrm{ch}}$ all non-local contributions are negligible, suppressing the non-local diagrams in the Hedin equation. Hence, the analysis of fluctuations will yield similar insights as the one of DMFT in the charge channel (App.~\ref{App:DMFT}). This is not the case for the bosonic spin propagator, which remains of almost constant magnitude throughout the cluster in the insulating phase, see App.~\ref{App:decay} for details.\\
 As a next step, we will study the Hedin equation in the spin representation in more detail to understand the suppression of the local diagrams and the rise of the bosonic nearest-neighbor diagram.

\subsection{Full fluctuation diagnostics}
We aim now at disentangling the real space contributions of spin fluctuations to the overall insulating self-energy for $U=6t, T=0.067t$. As discussed before, a positive difference $\Delta\Sigma>0$ of the self-energy for low frequencies is indicating the possible presence of a pole in $\Sigma$ for $\i\nu_n\rightarrow 0$, making this difference an interesting quantity for the diagnostics across the MIT \cite{Arzhang2020,Simkovic2020,Schaefer2021}. Hence we define the difference of the self-energy for the specific bosonic frequency $\i\omega_m$:
\begin{equation}
\Delta\tilde{\Sigma}(\i\omega_m)\coloneqq\tilde{\Sigma}_{\text{sp}}(\i\omega_m,\i\nu_{n=0})-\tilde{\Sigma}_{\text{sp}}(\i\omega_m,\i\nu_{n=1})
\end{equation}
for each \footnote{As discussed prior, the charge diagrams are purely local and their further analysis is, hence, obsolete.} real-space class of diagrams topologically possible between the sites of the $2\times2$ impurity.
Fig.~\ref{fig:FD_sigma_local} shows the real space fluctuation diagnostics of the difference of the self-energy  $\Delta\mathrm{Im}\tilde{\Sigma}_{\text{sp}}(\i\omega_m)$ in the left hand plot while the four right hand plots distinguish between different bosonic classes of diagrams for the case of $U=6t$.
For the bosonic propagator fixed to be local ($w_{\text{sp}}^\mathrm{00}$), all Feynman diagrams are given in Fig.~\ref{fig:FD_sigma_local}, panel $w^{00}$. The full contribution, arising from conducting the sum over all possible fermionic distances $f$ in the cluster is given in red square markers. We observe the contributions to $\Delta\Sigma$ from the local bosonic propagator to display a negative static peak and an equally as large positive peak for the first negative bosonic Matsubara frequency $\i\omega_{-1}$. The tag ``sum" in the bottom left hand corner gives the overall Matsubara sum over the bosonic fluctuations at the corresponding distances. This value is positive, and, hence, indicates a metallic behavior. The functional shape can be further decomposed according to the fermionic distance covered by the Green function in the Hedin equation. In the second panel of Fig.~\ref{fig:FD_sigma_local} we note, that for the local bosonic propagator, the positive contribution originates almost exclusively from the fermionic local diagram $G^\mathrm{00}$ (pink dashed line) while all diagrams together result in the static, insulating peak, exhibiting a comparatively small, metallic behavior when conducting the bosonic Matsubara $\text{sum}\!=\!0.9$. 
The overall negative, insulating contributions to the on-site self-energy originate from the bosonic nearest-neighbor diagrams $w^{10}$ and $w^{\overline{1}0}$, which are the same (within the Monte Carlo errors) due to cluster symmetry. Here the contribution of the local Green function diagram $G^\mathrm{00}$  (pink dashed line) is insulating (negative) for the lowest Matsubara frequencies. Remarkably, the leading summand is no longer the static bosonic frequency but rather the first negative one. The $G^\mathrm{02}$ contribution is only weak, while the left and right neighbor Green functions' ($G^\mathrm{01}$ and $G^\mathrm{0\overline{1}}$) contributions cancel each other (dashed dotted and dotted line, respectively). The bosonic second-nearest neighbor $w^\mathrm{20}$ contribution (rightmost panel) cancels in the bosonic sum and is hence negligible.

\begin{figure*}[t!]
    \centering
    \includegraphics[width=\linewidth]{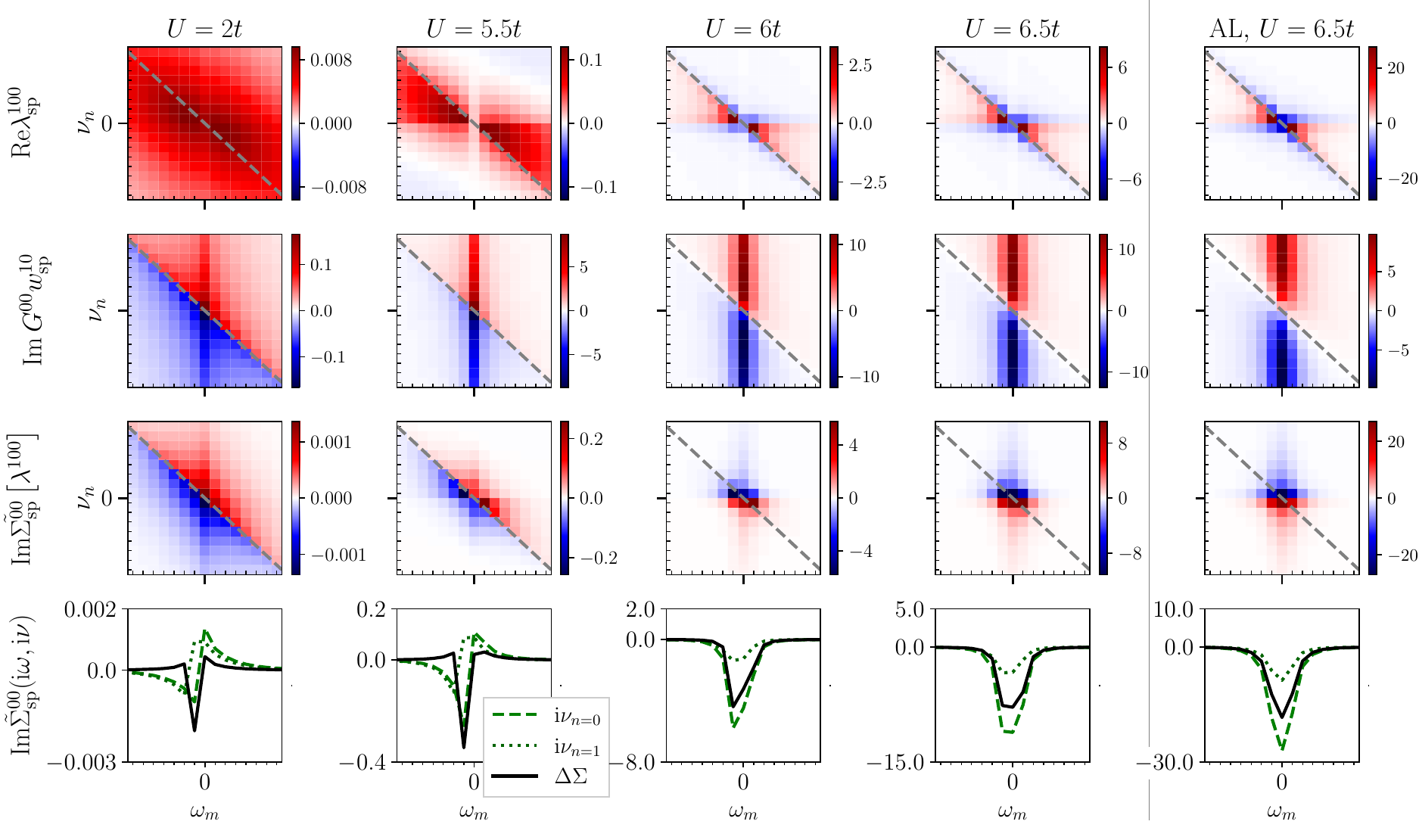}
    \caption{Analysis of the Hedin equation crossing the MIT. The first row shows the right neighbor Hedin vertex $\lambda$, while the second row displays the corresponding, imaginary part of the product of $G$ and $w$. The third row gives the product of $\lambda$ and $Gw$, i.e. their corresponding contribution to the local self-energy $\tilde{\Sigma}^\mathrm{00}\left[\lambda^{\mathrm{100}}\right](\mathrm{i}\omega_m,\mathrm{i}\nu_n)$. The fourth row gives cuts along the bosonic frequency of that quantity for fixed zeroth and first fermionic frequency (green dashed and dotted), respectively, and the resulting difference of the self-energy $\Delta\tilde{\Sigma}$ (solid line). Plots are sorted by column for increasing interaction values from left to right. On the very right-hand side, the corresponding result for the atomic limit (AL) cluster is shown. Grey dashed lines in the 2D plots indicate the antisymmetry of the Green function around $\i\omega+\i\nu=0$.}
    \label{fig:Hedin-to-sigma_local}
\end{figure*}

\subsection{Connecting the MIT to the structure of the Hedin vertex}
In this section, we aim at understanding first, how the nearest-neighbor spin-boson diagram leads to an insulating contribution to the self-energy (Sec.~\ref{sec:nn_hedin}), and, second, how the on-site spin-boson diagram gives an overall metallic contribution (Sec.~\ref{sec:loc_hedin}), in contrast to single-site DMFT.
\subsubsection{Why does the nearest-neighbor spin-boson diagram result in an insulating self-energy?}
\label{sec:nn_hedin}
As discussed before, the nearest-neighbor spin-boson contributions to the self-energy increase strongly on the insulating side of the MIT, specifically the on-site self-energy being dominated by the $w^\mathrm{10},G^\mathrm{00}$ diagram. This combination corresponds to the sketched Hedin equation of Fig.~\ref{fig:Hedin_example}. Here, the respective Hedin vertex is $\lambda_\text{sp}^\mathrm{100}\left(\i\omega_m,\i\nu_n\right)$, where the bosonic tip of the vertex corresponds to a nearest-neighboring distance to the fermionic ends, which are on the same site. In this subsection we further analyze the separate constituents of this diagram and illucidate the evolution of the Hedin vertex and its impact on the on-site self-energy across the Mott transition.

The Hedin vertex $\lambda_\text{sp}^\mathrm{100}\left(\i\omega_m,\i\nu_n\right)$, as a function of its frequencies, is displayed in the first row of Fig.~\ref{fig:Hedin-to-sigma_local} for increasing interaction values and $T=0.067t$. 
For the weak-coupling case $U=2t$, we observe a uniform broad positive structure. In the renormalized metallic regime ($U=5.5t$) the static frequency contributions are suppressed, while all contributions remain of positive sign. At the MIT ($U=6t$), first the static contributions change sign \cite{vanLoon2018} and, second, the Hedin vertex displays a node along the Fermi level $\i\omega+\i\nu=0$, highlighted by a dashed grey line. For larger interactions ($U\gtrsim 6.5t$), the frequency structure of the Hedin vertex approaches qualitatively the one of the isolated Hubbard cluster (AL), i.e., without an embedding into a bath.

Apart from the Hedin vertex, the second ingredient to the Hedin equation is the product
\begin{equation}
    Gw \coloneq G^{00}(i\omega_m+i\nu_n)w_\text{sp}^{10}(i\omega_m),
\end{equation}
whose imaginary part is shown in the second row of Fig.~\ref{fig:Hedin-to-sigma_local}. The dominant frequency feature of $\mathrm{Im }Gw$ is a change of sign of $Gw$ around the frequency $\i\omega+\i\nu=0$, indicated by a grey line in the plots. This nodal structure originates from the antisymmetry of the fermionic propagator $G$ and the symmetry of the bosonic one $w$ around their respective frequency arguments. Since in the weak coupling case of $U=2t$, the Hedin vertex is uniform, the nodal structure of $Gw$ is directly translated to the contribution to the self-energy:
\begin{equation}
\label{Eq:Contribution_to_sigma}
\begin{split}
\tilde{\Sigma}^\mathrm{00}_{\text{sp}}&\left[\lambda_\text{sp}^\mathrm{100}\right](\i\omega_m,\i\nu_n)\\&\coloneqq UG^\mathrm{00}\left(\i\omega_m+\i\nu_n\right)w^\mathrm{10}_{\text{sp}}\left(\i\omega_m\right)\lambda_{\text{sp}}^\mathrm{100}\left(\i\omega_m,\i\nu_n \right )\\&=UGw\lambda_{\text{sp}}^\mathrm{100}\left(\i\omega_m,\i\nu_n \right ),
\end{split}
\end{equation}
displayed in the third row of Fig.~\ref{fig:Hedin-to-sigma_local} as a function of its two Matsubara frequencies. This produces a metallic self-energy in the following way: For the first fermionic Matsubara frequency, the node in $\tilde{\Sigma}^\mathrm{00}_{\text{sp}}$ occurs between the $0^\text{th}$ and $1^\text{st}$ bosonic frequency $\omega_m$, resulting in an almost antisymmetric self-energy contribution in the third row (green dashed line). When performing the bosonic Matsubara sum this leads to an almost complete cancellation of contributions. This argument holds for the strongly renormalized metal at $U=5.5t$.

For the second fermionic Matsubara frequency, in contrast, the node in $\tilde{\Sigma}^\mathrm{00}_{\text{sp}}$ shifts to in between the $1^\text{st}$ and $2^\text{nd}$ bosonic frequency $\omega_m$. This leads to a reduced antisymmetry in the contribution to the self-energy (green dotted line), resulting in a weaker cancellation in the bosonic Matsubara sum than for the first fermionic Matsubara frequency and hence, to a metallic behavior (black solid line). We stress again, that this metallic self-energy is a manifestation of an (almost) structureless Hedin vertex and of the mismatch in bosonic excitation frequency and the pole in the Green function at the Fermi level, encoded as the change of sign at $\i\omega+\i\nu=0$. Hence it arises already from the weak coupling contribution in $U$ of the Hedin equation diagram.

The situation drastically changes in the insulating case ($U=6t$), where the Hedin vertex \textit{also} develops a node along $\i\omega+\i\nu=0$, compensating the node of $Gw$. Hence, $\mathrm{Im}\tilde{\Sigma}^{\mathrm
00}_\mathrm{sp}\left[\lambda^{\mathrm{100}}\right]$ is becoming a symmetric function without a node along $\i\omega_m+i\nu_n$, hence no cancellation occurs when conducting the bosonic frequency sum, leading to an overall large self-energy. 
Further, the Hedin vertex at static frequency $\i\omega_m=0$ changes sign when going from metallic to the insulating regime gradually, resulting in a suppressed self-energy contribution in this case. Hence, the maximum of $\tilde{\Sigma}^{\mathrm
00}_\mathrm{sp}$ occurs at the first negative bosonic frequency. For even larger interactions, at $U=6.5t$, the magnitude of the static part of the Hedin vertex increases again with respect to the MIT case, leading to a dominant zero-frequency term in the fluctuation diagnostics of the self-energy, which is qualitatively very similar to the atomic limit (last column), a detailed fluctuation diagnostics analysis of which can be found in App.~\ref{App:Hubbard_cluster}.
\begin{figure}[t!]
    \centering
    \includegraphics[width=\linewidth]{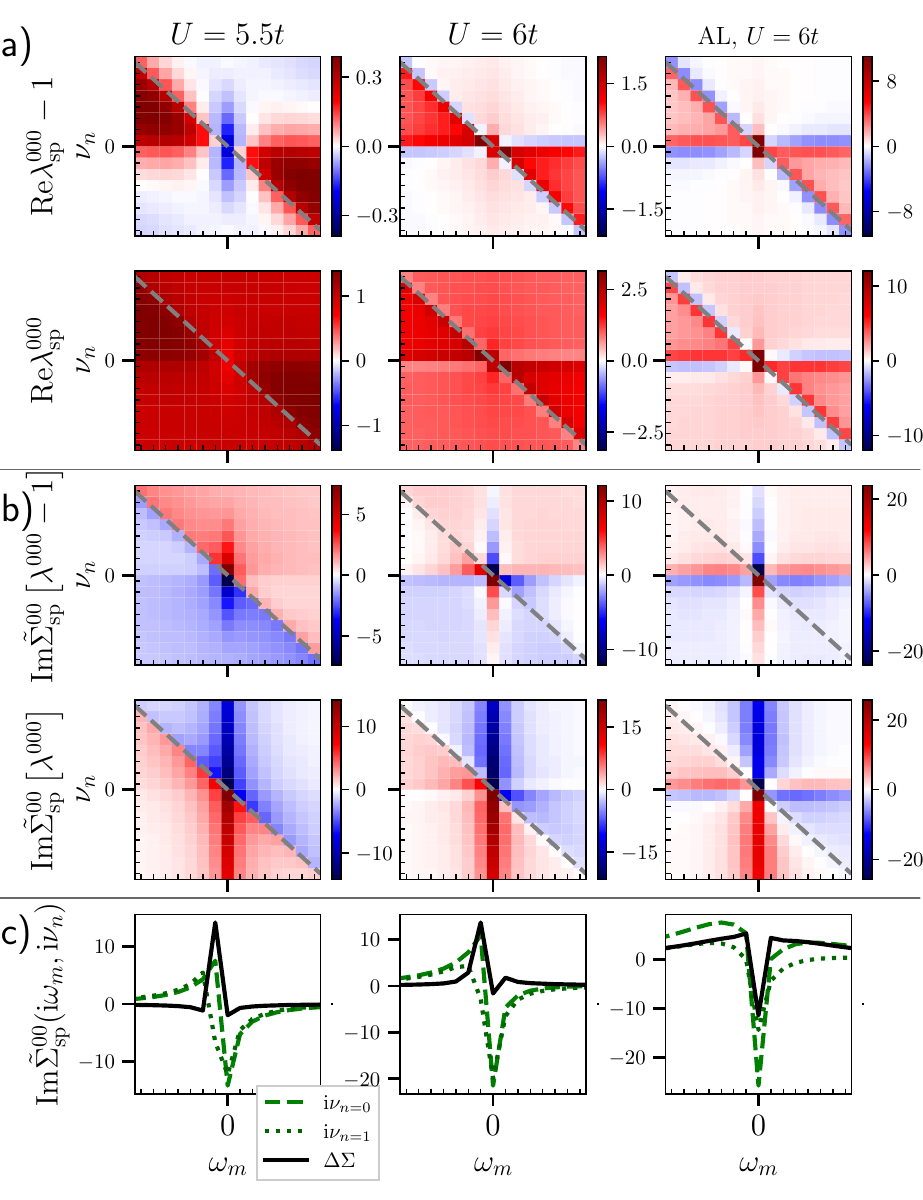}
    \caption{Analysis of the MIT via the Hedin vertex for interactions $U\!=\!5.5t$ (left column), $U\!=\!6t$ (central column), and AL (right column). a) contrasts the higher order contributions of the local Hedin vertex $\lambda$, after subtracting the first order perturbation theory contribution to the the full Hedin vertex. b) displays the corresponding contributions to the local self-energy term, first neglecting the first order contribution $\tilde{\Sigma}^\mathrm{00}\left[\lambda^{\mathrm{000}}-1\right](\mathrm{i}\omega_m,\mathrm{i}\nu_n)$ and then for the full Hedin vertex $\tilde{\Sigma}^\mathrm{00}\left[\lambda^{\mathrm{000}}\right](\mathrm{i}\omega_m,\mathrm{i}\nu_n)$. c) presents cuts along the bosonic frequency of the full contribution to the self-energy for the zeroth and first fermionic frequency (green dashed and dotted) and the resulting difference of the self-energy $\Delta\tilde{\Sigma}$ (solid line). Grey dashed lines in the 2D plots mark the symmetry of $G^{00}w^{00}$ around $\i\omega+\i\nu=0$.}
    \label{fig:Hedin_local-to-sigma_local}
\end{figure}

\begin{figure*}[t!]
    \centering
\begin{tikzpicture} 
\begin{feynman}
\vertex (i) {$\mathrm{0}$}; 
\vertex [left=4.05cm of i] (i0){ }; 
\vertex [right=.8cm of i] (n) {$f$}; 
\vertex [above right=0.8cm of n] (m) {$\mathrm{0}$};
\vertex [below right=0.8cm of m] (j) {$\mathrm{2}$};
\diagram* {(i) -- [fermion] (n) -- (j) -- (m)--(n),
(i) --[anti charged boson, quarter left](m),
}; 
\vertex [right=3.54cm of i] (i2) {$\mathrm{0}$}; 
\vertex [right=.8cm of i2] (n2) {$f$}; 
\vertex [above right=0.8cm of n2] (m2) {$\mathrm{1}$};
\vertex [below right=0.8cm of m2] (j2) {$\mathrm{2}$};
\diagram* {(i2) -- [fermion] (n2) -- (j2) -- (m2)--(n2),
(i2) --[anti charged boson, quarter left](m2),
}; 
\vertex [right=3.54cm of i2] (i3) {$\mathrm{0}$}; 
\vertex [right=.8cm of i3] (n3) {$f$}; 
\vertex [above right=0.8cm of n3] (m3) {$\mathrm{\overline{1}}$};
\vertex [below right=0.8cm of m3] (j3) {$\mathrm{2}$};
\diagram* {(i3) -- [fermion] (n3) -- (j3) -- (m3)--(n3),
(i3) --[anti charged boson, quarter left](m3),
}; 
\vertex [right=3.54cm of i3] (i4) {$\mathrm{0}$}; 
\vertex [right=.8cm of i4] (n4) {$f$}; 
\vertex [above right=0.8cm of n4] (m4) {$\mathrm{2}$};
\vertex [below right=0.8cm of m4] (j4) {$\mathrm{2}$};
\diagram* {(i4) -- [fermion] (n4) -- (j4) -- (m4)--(n4),
(i4) --[anti charged boson, quarter left](m4),
}; 
\end{feynman}
\end{tikzpicture}
    \includegraphics[width=1\linewidth]{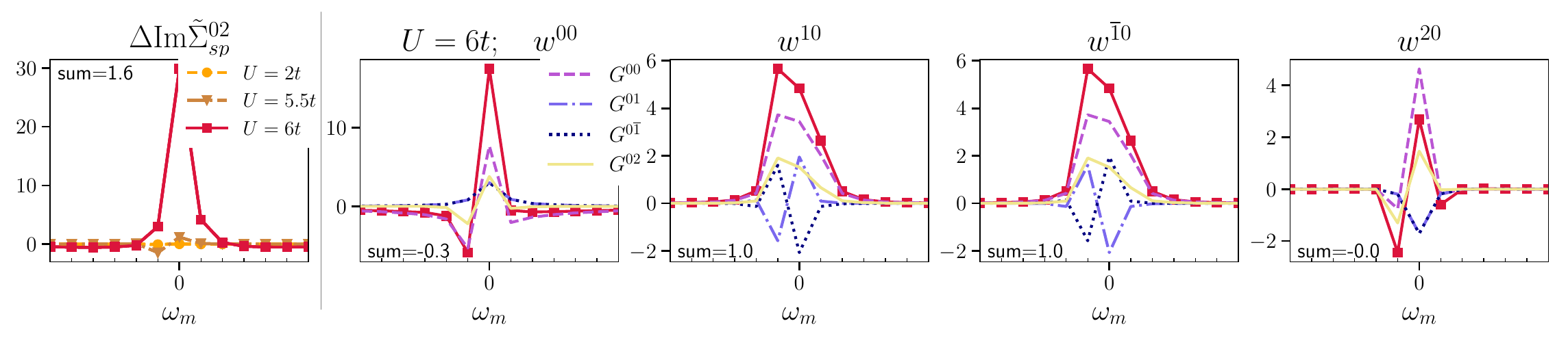}
    \caption{RFD in analogy to Fig.~\ref{fig:FD_sigma_local}, but for the imaginary part of the second-next neighbor self-energy.}
    \label{fig:FD_sigma_2nn}
\end{figure*}

\subsubsection{Why does the on-site spin-boson diagram result in a metallic contribution to the self-energy?}
\label{sec:loc_hedin}
While we have just discussed the origin of the insulating contribution from the nearest-neighbor spin-boson diagrams to the on-site self-energy in detail, we now discuss why the \textit{purely local} Hedin vertex $\lambda_\text{sp}^\mathrm{000}\left(\i\omega_m,\i\nu_n\right)$ does \textit{not} result in an insulating contribution to the on-site self-energy (in contrast to the results in single-site DMFT calculations). First, when recalling the fluctuation diagnostics of $\Delta\tilde{\Sigma}^{00}$ in Fig.~\ref{fig:FD_sigma_local}, we find that the diagram with $w_\mathrm{sp}^\mathrm{00}$ shows a strong insulating peak for the static frequency $\i\omega_m=0$. This negative contribution stems from a sum over all diagrams containing all possible fermionic propagators on the cluster. However, this peak is overcompensated by a large metallic peak at the first negative bosonic Matsubara frequency.

We now want to clarify, how the metallic and insulating peaks emerge from $\lambda^{\mathrm{000}}_{\mathrm{sp}}$. To that end, we to compare the contributions from the full on-site Hedin vertex to its higher order contributions. We recall that the on-site Hedin vertex contains a bare contribution of $1$, allowing us to separate the higher order contributions by
\begin{equation}
\begin{split}
\mathrm{Im}\tilde{\Sigma}_{\mathrm{sp}}^{\mathrm{00}}\left[\lambda^{000}\right]&=G^{00}w^{00}\lambda^{000}\\&=G^{00}w^{00} \left(\lambda^{000}-1\right)+G^{00}w^{00}.
\end{split}
\end{equation}
Please also note that $G^{00}w^{00}$ displays, again, a node around $\i\omega+\i\nu=0$. With the analysis of the data shown in Fig.~\ref{fig:Hedin_local-to-sigma_local}a), we want to make two points:\\
\begin{enumerate}
    \item[(i)] First, in the upper row we subtracted the lowest order contribution and present $\lambda^{\mathrm{000}}_{\mathrm{sp}}-1$. This reveals a complex higher-order behaviour over the MIT: For the metallic solution ($U=5.5t$), we find a negative response for the bosonic static frequencies (blue blob), which is bordered by a strong, blurred positive structure for the dynamic response. Going to the insulating solution ($U=6t$), a node develops for the Fermi level $\i\omega+\i\nu=0$ and furthermore, the static frequency contributions change sign. This remains similar in the atomic limit (AL, third column).
    \item[(ii)] Second, contrasting the full Hedin vertex (i.e., including the first order contribution of $1$, second row) to $\lambda^{\mathrm{000}}_{\mathrm{sp}}-1$, we see that the first order contribution completely dominates over the higher order structure, and only a strictly positive (red) structure remains.
\end{enumerate}
Fig.~\ref{fig:Hedin_local-to-sigma_local}b) shows the respective contributions to the self-energy. The contributions from the higher order correlations $\lambda^{\mathrm{000}}_{\mathrm{sp}}-1$ can be seen in the third row. Here, for $U=5.5t$ we find a node around $\i\omega+\i\nu=0$, stemming from the node in $G^{00}w^{00}$. For the insulating and atomic limit, this node is lifted by a change in symmetry of $\lambda^{\mathrm{000}}_{\mathrm{sp}}-1$.
In contrast, the full contribution of $\lambda^{\mathrm{000}}$ to the self-energy, i.e., including the constant offset (fourth row), we see that no change of symmetry occurs anymore. This fact, hence, results in a cancellation of contributions when performing the bosonic Matsubara sum for both, the  metallic and insulating case.

\section{Analyzing the non-local real-space self-energies}
\label{sec:sig_nonloc}
So far we have investigated the real-space fluctuations of which the metallic or insulating behavior of the on-site self-energy emerges. From the perspective of a momentum dependent self-energy (see Sec.~\ref{sec:discussion}) also non-local contributions stemming --on a $2\times 2$ cluster-- from nearest-neighbor and second-nearest-neighbor sites may have significant influence on the spectral properties. Hence, in analogy to the on-site case, we analyse these non-local components in the following subsections. We start with the second-nearest neighbor component (Sec.~\ref{sec:2nn}), as it overall displays a great similarity to the on-site case, before we address the nearest-neighbor component (Sec.~\ref{sec:nn}).

\subsection{Decomposing the second-nearest neighbor self-energy\texorpdfstring{ $\Sigma^\mathrm{02}_{\mathrm{sp}}(\mathrm i\nu_n)$}{}}
\label{sec:2nn}

We focus here exclusively on the spin representation of the second-next-neighbor self-energy \footnote{The second-next neighbor self-energy modulates the self-energy in the Brillouin zone as $\Sigma(\vec{k})\propto \Sigma^\mathrm{02}\cdot \left[\mathrm{cos}(k_x+k_y)+\mathrm{cos}(k_x-k_y)\right]$ \cite{Parcollet2004}. Its positive, imaginary pole of $\Sigma^\mathrm{02}$ acts thus as enlarging the imaginary part of the antinodal self-energy $\Sigma(0,\pi),\Sigma(\pi,0)$, while no modulation of the imaginary part of the nodal self energy takes place. Hence, we speak of an insulating contribution, acting in the antinode if the self-energy's difference is positive.}, as in the charge representation, the non-local diagrams do not contribute to the self-energy.
For the spin channel, a systematic decomposition of the difference of the self-energy $\Delta\tilde{\Sigma}_{\text{sp}}^\mathrm{02}$ via the Hedin equation is displayed in Fig.~\ref{fig:FD_sigma_2nn}. The only qualitative difference to the on-site case is that the second-nearest neighbor self-energy is \textit{positive} for positive fermionic Matsubara frequencies, instead of being \textit{negative}. 

Similar to the RFD of the on-site self-energy, a large positive, pole-like contribution originates from the bosonic nearest-neighbor diagrams $w^\mathrm{10}$ and $w^\mathrm{\overline{1}0}$ and simultaneously fermionically local diagrams $G^\mathrm{00}$. The discussion of the corresponding Hedin vertex $\lambda^{b,f,e}=\lambda_{\text{sp}}^\mathrm{102}$ is in complete analogy to the one of the on-site self-energy in Fig.~\ref{fig:Hedin-to-sigma_local}. Also, similar to the on-site self-energy, the bosonically local diagrams display a static contribution, which is overcompensated by higher frequency dynamic contributions, originating mostly from the $G^\mathrm{00}$ diagram. For the non-vanishing fermionic local diagram, the Hedin vertex $\lambda^\mathrm{002}$ (not shown) does not develop a clear node at $\i\omega+\i\nu=0$, and hence does not produce a significant contribution in the insulating case when performing the Matsubara sum.

\begin{figure*}[t!]
    \centering
    \begin{tikzpicture} 
\begin{feynman}
\vertex (i) {$\mathrm{0}$}; 
\vertex [left=4.2cm of i] (i0){ }; 
\vertex [right=.8cm of i] (n) {$f$}; 
\vertex [above right=0.8cm of n] (m) {$\mathrm{0}$};
\vertex [below right=0.8cm of m] (j) {$\mathrm{1}$};
\diagram* {(i) -- [fermion] (n) -- (j) -- (m)--(n),
(i) --[anti charged boson, quarter left](m),
}; 
\vertex [right=3.4cm of i] (i2) {$\mathrm{0}$}; 
\vertex [right=.8cm of i2] (n2) {$f$}; 
\vertex [above right=0.8cm of n2] (m2) {$\mathrm{1}$};
\vertex [below right=0.8cm of m2] (j2) {$\mathrm{1}$};
\diagram* {(i2) -- [fermion] (n2) -- (j2) -- (m2)--(n2),
(i2) --[anti charged boson, quarter left](m2),
}; 
\vertex [right=3.4cm of i2] (i3) {$\mathrm{0}$}; 
\vertex [right=.8cm of i3] (n3) {$f$}; 
\vertex [above right=0.8cm of n3] (m3) {$\mathrm{\overline{1}}$};
\vertex [below right=0.8cm of m3] (j3) {$\mathrm{1}$};
\diagram* {(i3) -- [fermion] (n3) -- (j3) -- (m3)--(n3),
(i3) --[anti charged boson, quarter left](m3),
}; 
\vertex [right=3.4cm of i3] (i4) {$\mathrm{0}$}; 
\vertex [right=.8cm of i4] (n4) {$f$}; 
\vertex [above right=0.8cm of n4] (m4) {$\mathrm{2}$};
\vertex [below right=0.8cm of m4] (j4) {$\mathrm{1}$};
\diagram* {(i4) -- [fermion] (n4) -- (j4) -- (m4)--(n4),
(i4) --[anti charged boson, quarter left](m4),
}; 
\end{feynman}
\end{tikzpicture}
    \includegraphics[width=1\linewidth]{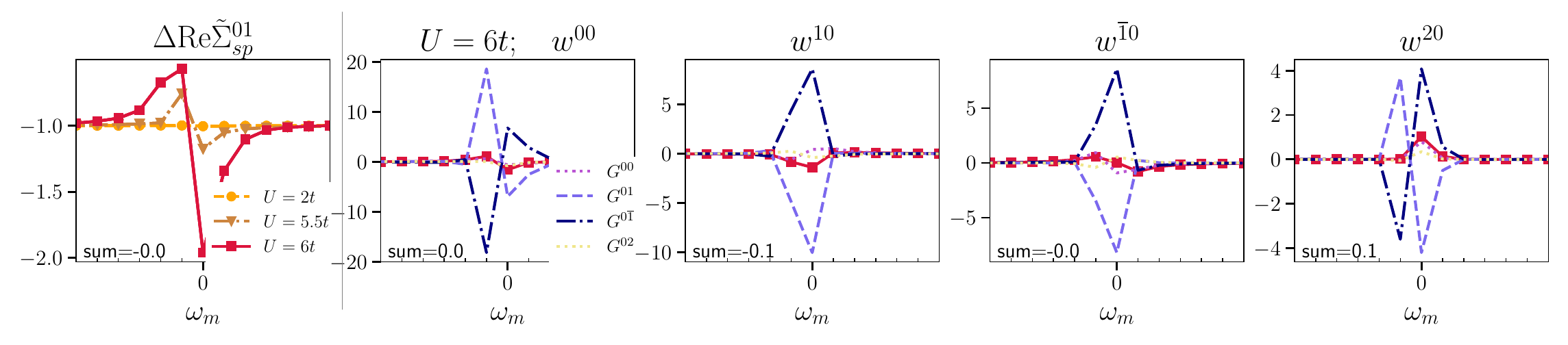}
    \caption{RFD in analogy to Fig.~\ref{fig:FD_sigma_local}, but for the real part of the next neighbor self-energy.}
    \label{fig:FD_sigma_n}
\end{figure*}

\subsection{Decomposing the nearest-neighbor self-energy\texorpdfstring{ $\Sigma^\mathrm{01}_{\mathrm{sp}}(\i\nu_n)$}{}} 
\label{sec:nn}
Eventually, we turn to the analysis of the nearest-neighbor self-energy $\Sigma^\mathrm{01}_{\text{sp}}(\i\nu_n)$. This part of the self-energy is fully real and acts as an effective modulation of the dispersion (see Sec.~\ref{sec:discussion}). Consistent with this property, the nearest-neighbor self-energy does not change qualitatively over the interaction-driven MIT, see left-most panel in Fig.~\ref{fig:FD_sigma_n}. As it can be deduced from the real-space fluctuation diagnostics in that figure, this is due to each diagram having a direct counterpart: the left-hand neighbor counters the right-hand neighbor fermionic propagator, and vice versa, in the Hedin equation. We will exemplify this for the next-neighbor boson part $w^\mathrm{10}$, which drives the insulating behavior for the other self-energies in the cluster: Here, the dashed and dashed-dotted lines, corresponding to the fermionic right- and left neighbor contributions $G^\mathrm{01},G^\mathrm{0\overline{1}}$ cancel almost completely, resulting in a strongly suppressed overall self-energy contribution (red solid line).

\begin{figure}[t!]
    \centering
    \begin{tikzpicture} 
\begin{feynman}
\vertex (i) {$\mathrm{0}$}; 
\vertex [left=.64cm of i] (i0){ }; 
\vertex [right=.8cm of i] (n) {$\mathrm{1}$}; 
\vertex [above right=0.8cm of n] (m) {$\mathrm{1}$};
\vertex [below right=0.8cm of m] (j) {$\mathrm{1}$};
\diagram* {(i) -- [fermion] (n) -- (j) -- (m)--(n),
(i) --[anti charged boson, quarter left](m),
}; 
\vertex [right=4.1cm of i] (i2) {$\mathrm{0}$}; 
\vertex [right=.8cm of i2] (n2) {$\mathrm{\overline{1}}$}; 
\vertex [above right=0.8cm of n2] (m2) {$\mathrm{1}$};
\vertex [below right=0.8cm of m2] (j2) {$\mathrm{1}$};
\diagram* {(i2) -- [fermion] (n2) -- (j2) -- (m2)--(n2),
(i2) --[anti charged boson, quarter left](m2),
}; 
\end{feynman}
\end{tikzpicture}
    \includegraphics[width=1\linewidth]{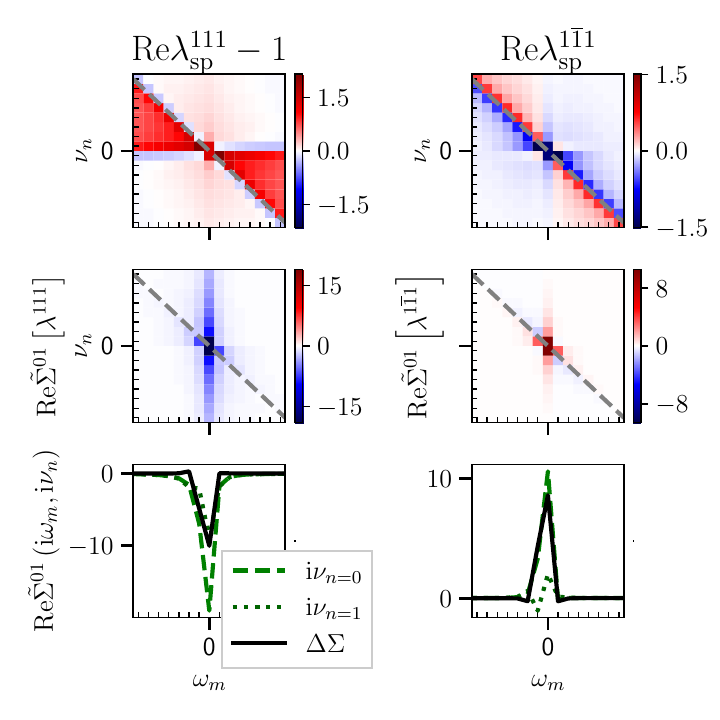}
    \caption{Two Hedin vertices at $U=6t$, $T=0.066t$ leading to a very similar (but opposite sign) contribution to the nearest-neighboring self-energy $\Sigma_\mathrm{sp}^\mathrm{01}$: The first row gives first the local Hedin vertex $\lambda^{\mathrm{111}}$ and the second row the corresponding contribution to the nearest-neighbor self-energy $\tilde{\Sigma}^\mathrm{01}_\mathrm{sp}(\mathrm{i}\omega_m,\mathrm{i}\nu_n)$ from the  Hedin vertex above. The third row gives cuts along the bosonic frequency of that quantity for fixed zeroth and first fermionic frequency (green dashed and dotted) and the resulting difference of the self-energy $\Delta\tilde{\Sigma}$ (solid line). In contrast, the second column shows the Hedin vertex $\lambda^{\mathrm{1\overline{1}1}}$ and subsequently, the corresponding contribution to the self-energy. The grey dashed line in the 2D plots indicate the point-symmetry of the Green function around $\i\omega+\i\nu=0$.}
    \label{fig:Compare_Sig_nn}
\end{figure}

We aim now at connecting the cancellation of the self-energy $\Sigma^{01}_{\mathrm{sp}}$ with the two involved Hedin vertices, (i) $\lambda^{b,f,e}=\lambda^\mathrm{111}_\mathrm{sp}$ and (ii) $\lambda^{b,f,e}=\lambda^\mathrm{1\overline{1}1}_\mathrm{sp}$, respectively: 

(i) For the diagram where the fermionic propagator corresponds to a right-neighbor distance $G^\mathrm{01}$, the first column of Fig.~\ref{fig:Compare_Sig_nn} displays the associated, completely local Hedin vertex $\lambda^\mathrm{111}_\mathrm{sp}-1$, where we subtracted the first order contribution. We can see from the second and third row of Fig.~\ref{fig:Compare_Sig_nn}, that only the zero frequency $\i\omega_m=0$ contributes to $\Delta\Sigma$, while the higher order features are suppressed, reasoned by a sharp bosonic propagator. Note, that the first order constant contribution of $1$ does not alter $\Delta\Sigma$.

(ii) For the left-neighbor fermionic propagator $G^\mathrm{0\overline{1}}$, the Hedin vertex $\lambda^\mathrm{1\overline{1}1}_\mathrm{sp}$, where one fermionic leg corresponds to a second-next neighbour distance, is displayed in the second column of Fig.~\ref{fig:Compare_Sig_nn}. We also find a node for $\i\omega+\i\nu=0$ apart from for $\i\omega_0$. We observe that the static frequency contributions are actually of opposite sign w.r.t. the one of $\lambda^\mathrm{111}_\mathrm{sp}-1$. Further, the self-energy $\tilde{\Sigma}^{\mathrm{01}}_\mathrm{sp}\left[\lambda^\mathrm{1\overline{1}1}\right]$ only displays significant contributions for the static frequency. These are of opposite sign w.r.t. the ones in the first column. This results then in a self-energy's difference of very similar magnitude but with opposite sign.

The change of sign in the Hedin vertex can be explained from the Green function diagrams contained: The Green function, with which the diagrams in the Hedin vertex are multiplied, is picking up a factor $\i$ for each nearest-neighboring distance it covers on the square lattice at half-filling \footnote{This can already be seen from the (non-interacting) Green function of the Hubbard dimer, where the diagonal elements are complex and the off-diagonal elements are real and negative. In the Lehmann representation of the Green function, this is a field-theoretic manifestation of the Jordan-Wigner transformation and operator ordering on the Fock space to preserve the commutation relations for the fermionic creation and annihilation operators.}. Hence, covering a second-nearest neighboring distance corresponds to a prefactor of $\i^2=-1$. In fact, when analyzing the higher order contributions of the local Hedin vertex $\lambda_{\text{sp}}^\mathrm{111}$, i.e., neglecting the first order contribution of $-1$, we find a qualitatively similar frequency structure as for $\lambda_{\text{sp}}^\mathrm{1\overline{1}1}$, but of opposite sign (not shown). This symmetry argument applies to all contributions to the nearest-neighboring self-energy and, therefore, diagrams from the right and left neighboring fermionic propagators (e.g., $G^{01}$ and $G^{0\overline{1}}$) tend to overall cancellation.

\section{Summary, discussion, and conclusions}
\label{sec:conclusion}
In this last section, after summarizing our main results in real space in Sec.~\ref{sec:summary}, we will discuss their implications on momentum-space representations of correlations in Sec.~\ref{sec:discussion}, before concluding and giving an outlook in Sec.~\ref{sec:outlook}.

\subsection{Summary of our results in real space}
\label{sec:summary}
The description of the self-energy via a low-frequency spin picture allows to precisely identify the local and non-local fluctuations which drive the metal-insulator transition. While the real-space decomposition of the Hedin equation for the local (on-site) self-energy demonstrates that contributions stemming from the local bosonic spin propagator become more metallic over the MIT, the non-local spin fluctuations on the $N_c=2\times 2$ CDMFT cluster produce an insulating local self-energy, reducing the critical interaction relative to $N_c=1$ DMFT. This statement remains true when regarding non-local single-boson exchange spin diagrams \cite{Yu2024}, see App.~\ref{App:SBE}. Specifically, we found that the enhancement of the nearest-neighbor spin contribution, yielding by far the largest contributions to all real-space self-energies, is directly related to a sudden change in the symmetry of the Hedin vertex at the MIT. First, this vertex abruptly develops a node around the Fermi surface $\i\omega+\i\nu=0$ and $\i\nu=0$ for non-zero bosonic frequencies $\i\omega\neq0$, leading to enhanced contributions to the self-energy. Second, the fact that we find a smooth change of sign only in the non-local Hedin vertex for the zero frequencies $\i\omega=0$ represents an important difference with respect to the findings of DMFT, where the local Hedin vertex displays the change of sign \cite{vanLoon2018} at the MIT, see App.~\ref{App:DMFT}.

\begin{figure}[t!]
    \centering
    \includegraphics[width=\linewidth]{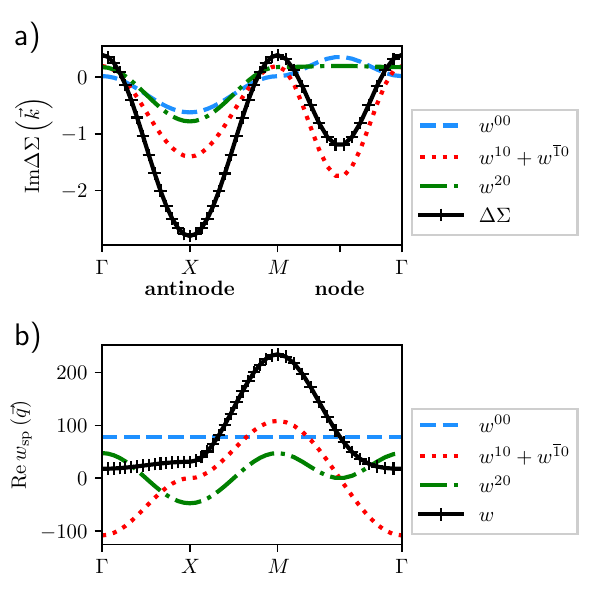}
    \caption{a) The contributions to the difference of the self-energy $\Delta\Sigma(\vec{k})$ along a path in the Brillouin zone, originating from the bosonic propagator computed at different distances within the cluster for the insulating case of $U=6.5t$ and $T=0.067t$. Contributions from different disjoint classes of diagrams contain either the local spin-boson $w^\mathrm{00}$ (blue dashed line), the nearest-neighbor spin-boson $w^\mathrm{01}+w^\mathrm{0\overline{1}}$ (red dotted line) or the second-nearest neighbor spin-boson $w^\mathrm{20}$ (green dot-dashed line). The three summands add up to the total difference of the self-energy (black bold line). The antinode and node are given along the $\vec{k}$-axis. b) The different contributions of the real-space spin-boson propagator to a periodized, static spin-boson propagator $w(\vec{q})$ (black bold line) along a path in the Brillouin zone for the same parameters as in panel a).}
    \label{fig:SigSp(k)}
\end{figure}

\subsection{Discussion on the impact on momentum space}
\label{sec:discussion}
We aim to set our real-space investigation into context by discussing the resulting momentum-dependent self-energy, periodized via a Fourier transform \cite{Parcollet2004}. We note, first, that the on-site self-energy evidently results in a constant offset of all momentum-space quantities and, therefore, corresponds to an average over the Brillouin zone, while non-local real space self-energies modulate the quantity over the Brillouin zone.

The black bold line in Fig.~\ref{fig:SigSp(k)} a) displays the difference $\Delta\Sigma$ of the self-energy along a high-symmetry path in the Brillouin zone while the other lines display the contributions from certain spin-boson propagators in real space. This path includes the antinode [$\vec{k}_\text{AN}\!=\!X\!=\!(\pi,0)$] as well as the node [$\vec{k}_\text{N}\!=\!(\pi/2,\pi/2)$]. We note, that the nodal and antinodal contributions are mostly governed by the nearest-neighboring spin-boson propagator $w^\mathrm{10}$ (red dotted line). In particular, while in the antinodal direction, both the local $w^\mathrm{00}$ (blue dashed line) and second-nearest neighbor spin-boson propagator $w^\mathrm{20}$ (green dashed line) further enhance the insulating nature of the self-energy originating from $w^\mathrm{10}$ (i.e., they make $\Delta\Sigma$ more negative), they are zero or even metallic at the node, leading to an enhanced nodal - antinodal dichotomy.  Let us also note that neglecting the contribution of the nearest-neighboring spin-boson $w^\mathrm{10}$ [red dotted line in Fig.~\ref{fig:SigSp(k)} a)], would result in an insulating self-energy at the antinode and a metallic self-energy at the nodal point. For the remaining contributions $w^\mathrm{00},w^\mathrm{20}$, we find, similar to \cite{Krien2022}, that the zero frequency contribution is dominating in the half-filled case.

The analysis of the periodized spin propagator $w_\text{sp}(\vec{q})$ itself, shown in Fig.~\ref{fig:SigSp(k)} b), displays an antiferromagnetic peak at $\vec{q}\!=\!M\!=\!(\pi,\pi)$, where also the next-neighbour spin-boson propagator $w^\mathrm{10}$ (red dotted line), principal contributor to the self-energy, causes the largest contribution to the spin propagator. This illustrates the connection between the insulating self-energy and the pronounced antiferromagnetic peak, further supporting the findings of a strong impact of AF fluctuations in \cite{Gunnarsson2015,Gunnarsson2018,Bonetti2022,Arzhang2020} on the system. Our results, however, shed also light on the precise nature of the underlying physical processes: \textit{dynamic} low-frequency spin fluctuations over the next-neighbor distance drive the MIT in cluster-DMFT.

Consistently with this picture, we find that the bosonic nearest-neighbor contribution, driving the AF fluctuations, is negative at the $\Gamma$-point, which decreases the ferromagnetic correlations and vanishes for the $X$-point, which corresponds to stripe-spin order \footnote{Please note that the shape of $w^{01}(\vec{q})$ is given by the dispersion relation, since we only consider nearest-neighbor hopping.}.
\subsection{Conclusions and outlook}
\label{sec:outlook}
In this paper we introduced the real-space fluctuation diagnostics technique, and presented a first application for the half-filled Hubbard model on a square lattice. By analyzing its CDMFT solution, we demonstrated that nearest-neighbor, low-frequency \textit{dynamic} antiferromagnetic spin-boson excitations are responsible for the occurrence of the MIT, see Fig.~\ref{fig:FD_sigma_local}. We achieved this finding by making a clear-cut connection of bosonic fluctuations with the single-particle fermionic self-energy. Concomitantly, our analysis further clarifies the crucial impact of non-perturbative, non-local features of the fermion-boson coupling vertex in the description of the MIT. Our surprising result may be the consequence of the particle-hole symmetry of the system. 
In fact, in recent studies of the weakly hole-doped pseudogap in the Hubbard model \cite{Gunnarsson2015,Gunnarsson2018,Krien2022}, \textit{static} non-local correlations were found to enhance the self-energy at the antinode and diminish the self-energy at the nodal point. In future studies, one could relax the condition of  particle-hole symmetry, by performing similar calculations in the doped regime of the Hubbard model. These studies could clarify numerically the role of resonating valence bond (RVB) fluctuations in opening the pseudogap \cite{Anderson87, Gunnarsson2018}.

\acknowledgments 
We would like to thank M\'ario Malcolms de Oliveira, Friedrich Krien, Kilian Fraboulet, Dominik Kiese, Henri Menke, Miriam Patricolo, and Demetrio Vilardi for fruitful discussions. The authors acknowledge the members of the computer service facility of the MPI-FKF for their help and C. Museum for intriguing input. We acknowledge financial support from the Deutsche Forschungsgemeinschaft (DFG)
within the research unit FOR 5413/1 (Grant No. 465199066). This research was also funded
by the Austrian Science Fund (FWF) 10.55776/I6946, A.T. acknowledges support from the Austrian Science Fund (FWF) through the grant 10.55776/I5868 (Project P1 of the research unit QUAST, for5249, of the German Research Foundation, DFG) and P.M.B. acknowledges support by the German National Academy of Sciences Leopoldina through Grant No. LPDS 2023-06 and funding from U.S. National Science Foundation grant No. DMR2245246 and the Gordon and Betty Moore Foundation’s EPiQS Initiative Grant GBMF8683. The Flatiron Institute is a division of the Simons Foundation.
\appendix
\section*{Appendices}
\section{Definitions, derivations, error estimate and CDMFT cycle}\label{App:Def}
This section first presents a derivation of the Hedin equation in the time-contour formulation in App.~\ref{App:Derivation1}, translate it to the real-space in App.~\ref{App:Derivation} and subsequently make the relevant connection to the Hedin vertex and the computed quantities in App.~\ref{App:Definition}.
\subsection{Derivation of the Hedin equation for bosonic fluctuations for the contour case}\label{App:Derivation1}
The local Hubbard interaction can be written as a Fierz decomposition \cite{Ayral2017,Krien2019_2}:
\begin{equation}
H_{\text{int}}=Un_{\uparrow}n_{\downarrow}=U\frac{rnn+(r-1)mm}{2}-\left(r-\frac{1}{2}\right)Un,
\end{equation} where $n=n_\uparrow+n_\downarrow$, $m=n_\uparrow-n_\downarrow$ and $r\in\left[0,1\right]$. This decomposition was used to reformulate the self-energy for the local (Anderson model) case \cite{Krien2019_2} and momentum space case \cite{Kiese2024} into spin and charge channels. We aim at obtaining the same result for the real-space case. In this section, we derive the self-energy in its decoupled form, mostly following \cite{Kugler} Eq.~(2.11) and \cite{Kopietz2010}, starting form Schwinger's action principle. We write the action as quartic expression, meaning we write $nn=n_{1'}\cdot n_{2'}=\overline{c}_{1'}c_{1'}\overline{c}_{2'}c_{2'}$ and switch for this section to a compact index notation $1',2'$ with $1'=(\tau_{1'},r_{1'})$ as tuple of real-space and imaginary time contour variables. We abbreviate the Hubbard interaction as $\Gamma^0_{1',2'}=U\delta_{r_{1'},r_{2'}}\delta(\tau_{1'}-\tau_{2'})$ to be local in space and time \cite{Baier2004}. Primed indices indicate internal indices, while non-primed indices are later external variables. Further, $\sigma, \sigma'\in\{\uparrow,\downarrow\}$ denote spin flavour, such that the action reads:
\begin{equation}
\begin{split}
S\left[\overline{c},c\right]=&-\sum_{\sigma}\int_{1',2'}\overline{c}_{1',\sigma}\left(G_0^{-1}\right)_{1',2'}c_{2',\sigma}\\
&+\int_{1',2'}\frac{r\Gamma^0_{1',2'}}{2}\left[(n_{1'\uparrow}+n_{1'\downarrow})(n_{2'\uparrow}+n_{2'\downarrow})\right]\\
&+\int_{1',2'}\frac{(r-1)\Gamma^0_{1',2'}}{2}\left[(n_{1'\uparrow}-n_{1'\downarrow})(n_{2'\uparrow}-n_{2'\downarrow})\right]\\
&-\left(r-\frac{1}{2}\right)\int_{1',2'}(n_{1'\uparrow}+n_{1'\downarrow})\Gamma^0_{1',2'}\,\\
\end{split}
\end{equation} and as preliminary for the further derivation:
\begin{equation}
\begin{split}
\frac{\delta S}{\delta \overline{c}_{1\uparrow}}=&-\int_{2'}\left(G_0^{-1}\right)_{1,2'}c_{2',\uparrow}\\
&+r\int_{2'}\Gamma^0_{1,2'}\left[c_{1\uparrow}(n_{2'\uparrow}+n_{2'\downarrow})\right]\\
&+(r-1)\int_{2'}\Gamma^0_{1,2'}\left[c_{1\uparrow}(n_{2'\uparrow}-n_{2'\downarrow})\right]\\
&-\left(r-\frac{1}{2}\right)\int_{2'}c_{1\uparrow}\Gamma^0_{1,2'}.\\
\end{split}
\end{equation}
 The Grassman integral is invariant under a (linear \footnote{Consider, that functions of Grassmann variables can only depend on up to first order terms, due to Pauli principle.}) shift of variables \cite{Kopietz2010} hence, the first order expansion term of the integrand must vanish:
\begin{equation}
\begin{split}
\label{Eq:Func_Derivative}
    0=&\int \mathcal{D}\left[ \overline{c},c\right]\frac{\delta}{\delta \overline{c}_{1\uparrow}}e^{-S+S_j},
\end{split}
\end{equation} where we introduce fermionic sources $\overline{j}_{1'},j_{1'}$:
\begin{equation}
    S_j=\sum_{\sigma}\overline{j}_{1'\sigma} c_{1'\sigma}+\overline{c}_{1'\sigma}j_{1'\sigma}.
\end{equation} Employing the chain rule in Eq.~(\ref{Eq:Func_Derivative}) and further, conducting the functional derivative $\frac{ \delta }{\delta j_{3'\uparrow}}$, we obtain:
\begin{equation}
\begin{split}
    0=\int_{2'}\int \mathcal{D}&\left[ \overline{c},c\right]\left[ \delta_{1,3'}+\left(G_0^{-1}\right)_{1,2'}c_{2',\uparrow}\overline{c}_{3',\uparrow}\right.\\
    &-\Gamma^0_{1,2'}\,r(c_{1\uparrow}\overline{c}_{3'\uparrow}n_{2'\uparrow}+c_{1\uparrow}\overline{c}_{3'\uparrow}n_{2'\downarrow}\\
    &-\Gamma^0_{1,2'}\,(r-1)(c_{1\uparrow}\overline{c}_{3'\uparrow}n_{2'\uparrow}-c_{1\uparrow}\overline{c}_{3'\uparrow}n_{2'\downarrow})\\
    &+\left.\left(r-\frac{1}{2}\right)\Gamma^0_{1,2'}\,c_{1\uparrow}\overline{c}_{3'\uparrow} \right]e^{-S+S_j},
\end{split}
\end{equation} We now set the sources to zero, divide by the partition function $\mathcal{Z}$, and eliminate the integral over $2'$ by making use of the delta-function in $\Gamma^0_{1',2'}=U\delta_{r_{1'},r_{2'}}\delta(\tau_{1'}-\tau_{2'})$, switching to operator formalism and introducing the fermionic propagator $G_{2',3'}=-\langle c_{2'\uparrow}c^\dagger_{3'\uparrow}\rangle=-\frac{1}{\mathcal{Z}}\int\mathcal{D}\left[\overline{c},c\right]c_{2'\uparrow}\overline{c}_{3'\uparrow}e^{-S}$:
\begin{equation}
\begin{split}
   \left(G_0^{-1}\right)_{1,2'}G_{2',3'}&-\delta_{1,3'}=\\
   &+rU\underbrace{\left[\langle c^{\dagger}_{3'\uparrow} c_{1\uparrow} n_{1\uparrow}\rangle+\langle c^\dagger_{3'\uparrow}c_{1\uparrow}n_{1\downarrow}\rangle\right]}_{\chi^3_{\mathrm{ch}}+nG}\\
   &+(r-1)U\underbrace{\left[\langle c^{\dagger}_{3'\uparrow}c_{1\uparrow}n_{1\uparrow}\rangle-\langle c^\dagger_{3'\uparrow}c_{1\uparrow}n_{1\downarrow}\rangle\right]}_{\chi^{3}_{\mathrm{sp}}}\\
   & +\left(\frac{1}{2}-r\right)UG_{1,3'}
\end{split}
\end{equation} where repeated indices are to be integrated over. Introducing the connected fermion-boson response function in the specific channel:
\begin{equation}
\begin{split}
\left(G_0^{-1}\right)_{1,2'}G_{2',3'}-\delta_{1,3'}=&(r-1)U\chi^3_{\mathrm{sp},1,1,3'}\\ 
&+rU\left[\chi^3_{\mathrm{ch},1,1,3'}+nG_{1,3'}\right]\\
   & +\left(\frac{1}{2}-r\right)UG_{1,3'}, 
\end{split}
\end{equation} multiplication from the right by $G^{-1}_{3',2}$ and employing the Dyson equation yields
\begin{equation}
\label{eq:fullhedin}
\begin{split}
    \Sigma_{1,2}=&(r-1)U\chi^3_{\mathrm{sp},1,1,3'}G^{-1}_{3',2}+rU\chi^3_{\mathrm{ch},1,1,3'}G^{-1}_{3',2}\\
    &+U\left((n-1)r+\frac{1}{2}\right)\delta_{1,2}\\
    \end{split}
\end{equation}
or writing the contour integrals explicitly:
\begin{equation}
\begin{split}
\label{Eq:Hedin_contour}
    \Sigma(1,2)=&(r-1)U\int d3'\,\chi^3_{\mathrm{sp}}(1,1,3')G^{-1}(3',2)\\
    &+rU\int d3'\, \chi^3_{\mathrm{ch}}(1,1,3')G^{-1}(3',2)\\
    &+\delta(1,2)\Sigma_C,
\end{split}
\end{equation}
where we denote the local part of the self-energy in space and time as $\Sigma_{C}=U\left((n-1)r+\frac{1}{2}\right)$.
As we now have a general contour formulation of the self-energy, we can continue Fourier transforming this result to the Matsubara description of a real-space lattice.
\subsection{Derivation of the Hedin equation in the Matsubara formalism for the lattice/orbital case}
\label{App:Derivation}
We write the fermion-boson response function in terms of the bosonic legs \begin{equation}w_{\text{sp/ch}}=U\pm U^2\chi_{\text{sp/ch}},\end{equation} the fermionic legs $G$ and the fermion-boson scattering amplitude (Hedin vertex) $\lambda_{\text{sp/ch}}$ \cite{Hedin1965}:
\begin{equation}
\begin{split}
    \chi_{\text{sp/ch}}^3(1,2,3)=-\int &d1'd2'd3'\, w_{\text{sp/ch}}(1;1')/U\\
    &\cdot\lambda_{\text{sp/ch}}(1';2';3') G(3';3)G(2';2)
\end{split}
\end{equation} and obtain for the self-energy Eq.~(\ref{Eq:Hedin_contour}) for $r=0$ (sp) and $r=1$ (ch), respectively:
 \begin{equation}
 \begin{split}
     \Sigma(1;2)=&\Sigma_C\delta(1,2)\\ &\pm\int d34 G(3;1)w_{\mathrm{sp/ch}}(1;4)\lambda_{\mathrm{sp/ch}}(4;2;3)
\end{split}
 \end{equation} where $1,2,3,4$ are combined space time variables. Note, that this finding differs from \cite{Biermann2008} only by the modification in the local part $\Sigma_{C}$. The $\pm$-sign takes into account that the effective interactions as the first order term in charge- and spin channels are of opposite sign \cite{Biermann2008,Ayral2015}. Making the space-time variables explicit as $1=(\tau_1,r_1)$ et.c., where $r$ is a respective lattice site index and $\tau$ the respective imaginary time index:
 \begin{equation}
 \label{Eq:InitialHedin}
 \begin{split}
     \Sigma^{r_1,r_2}(\tau_1,\tau_2)=&\Sigma_C\\ \pm &\sum_{r_3,r_4}\int_0^\beta \mathrm{d}\tau_3\mathrm{d}\tau_4 G^{r_3,r_1}(\tau_3,\tau_1)w^{r_1,r_4}_{\mathrm{sp/ch}}(\tau_1,\tau_4)\\&\qquad\cdot \lambda^{r_4,r_2,r_3}_{\mathrm{sp/ch}}(\tau_4,\tau_2,\tau_3).
\end{split}
 \end{equation}
 Introducing the imaginary time Fourier transform
 \begin{equation}
     \begin{split}
       G^{s,f}(\tau_3,\tau_1)=&-\langle T_\tau c_s(\tau_3){c}_f^{\dagger}(\tau_1)\rangle\\=&\frac{1}{\beta}\sum_m G^{s,f}(\i\nu_m)\mathrm{e}^{-\i\nu_m\left(\tau_1-\tau_3\right)},  \\
       w^{b,s}(\tau_1,\tau_4)=&U\delta^{b,s}\pm U^2\left[\langle T_\tau n_f(\tau_1)n_s(\tau_4)\rangle-n_f n_s\right]\\=&\frac{1}{\beta}\sum_k w^{b,s}(\i\omega_k)\mathrm{e}^{-\i\omega_k\left(\tau_4-\tau_1\right)}\\
       \lambda^{b,f,e}(\tau_4,\tau_2,\tau_3)=&\frac{1}{\beta^2}\sum_{l,d} \lambda^{b,f,e}(\i\omega_l,\i\nu_d)\mathrm{e}^{-\i\omega_l\left(\tau_3-\tau_4\right)-\i\nu_d\left(\tau_3-\tau_2\right)}\\
     \end{split}
 \end{equation} with orbital indices $s,f,b,e$, fermionic Matsubara \cite{Matsubara1955} frequencies $\i\nu$ and bosonic Matsubara frequencies $\i\omega$. The diagramatically incoming leg is always denoted by the last time variable.
 Inserting in Eq.~(\ref{Eq:InitialHedin}) and using
 \begin{equation}
     \int_0^{\beta}\mathrm{d}\tau\mathrm{e}^{-\i\omega_c\tau}=\beta\delta(\i\omega_c)
 \end{equation} for bosonic combinations of Matsubara frequencies $\i\omega_c$, one finds
\begin{equation}
\begin{split}
    &\Sigma^{s,e}(\tau_2,\tau_1)-\Sigma_C\\
    &=\pm\frac{1}{\beta^2}\sum_{b,f}\sum_{mkld}G^{s,f}(\i\nu_m)w^{b,s}(\i\omega_k)\lambda^{b,f,e}(\i\omega_l,\i\nu_d)\\&\qquad\cdot\delta(\i\omega_k-\i\omega_l)\delta(\i\nu_m-\i\nu_d-\i\omega_k)\mathrm{e}^{\i\tau_1(\omega_k-\nu_m)+\i\tau_2\nu_d}\\
    &=\pm\frac{1}{\beta^2}\sum_{b,f}\sum_{mk}G^{s,f}(\i\nu_m)w^{b,s}(\i\omega_k)\lambda^{b,f,e}(\i\omega_k,\i\nu_m-\i\omega_k)
    \\&\qquad\qquad\cdot\mathrm{e}^{\i(\nu_m-\omega_k)(\tau_2-\tau_1)}.\\
\end{split}
\end{equation}
Fourier transforming the self-energy and using time-translation invariance $\tau_1=0$:
\begin{equation}
    \begin{split}
        \Sigma^{s,e}&(\i\nu_n)-\Sigma_C=\pm\int_0^\beta\mathrm{d\tau_2}\mathrm{e}^{-\i\nu_n\tau_2}\Sigma^{s,e}(\tau_2,0)\\
        &=\pm\frac{1}{\beta}\sum_{d,f}\sum_{mk}G^{s,f}(\i\nu_m)w^{b,s}(\i\omega_k)\lambda^{b,f,e}(\i\omega_k,\i\nu_m-\i\omega_k)\\&\qquad\qquad\cdot\delta\left(\nu_m-\nu_n-\omega_k\right)\\
        &=\pm\frac{1}{\beta}\sum_{d,f}\sum_{k}G^{s,f}(\i\nu_n+\i\omega_k)w^{b,s}(\i\omega_k)\lambda^{b,f,e}(\i\omega_k,\i\nu_n),
        \end{split}
\end{equation} which resembles Eq.~(\ref{eq:Hedin}). Restricting the equation to a single site, we reproduce \cite{Krien2019_2}, note however the different conventions in the definition of the bosonic leg $w$.
We close this section with the remark, that in the particle-hole symmetric case, $\Sigma_C=U/2$ for both channels.
\subsection{The Hedin vertex and quantities as they are computed}\label{App:Definition}
We compute the fermi-bose response with the continuous-time quantum Monte Carlo (CT-INT, \cite{Maier2005,Rubtsov2005}) impurity solver, implemented using the TRIQS library \cite{TRIQS}. In the calculations we subtract the disconnected contributions:
\begin{equation}
\begin{split}
\chi^{3,\mathrm{b,f,e}}_{\text{ph},\sigma\sigma'}(\i\omega_m,\i\nu_n)=\int_0^\beta& \mathrm{d}\tau_2 \mathrm{d}\tau_3 \mathrm{d}\tau_4 \mathrm{e}^{-\i\nu_n(\tau_2-\tau_3)} \mathrm{e}^{-\i\omega_m(\tau_4-\tau_3)} \\&\left[\langle T_{\tau}(c^\mathrm{e})^\dagger_{\sigma} (\tau_3)c^\mathrm{f}_{\sigma}(\tau_2)n^\mathrm{b}_{\sigma'}(\tau_4)\rangle \right]\\&-n^{\mathrm{b}}_{\sigma'}\delta(\i\omega)G^{\mathrm{f,e}}_{\sigma}(\i\nu_n) ,
\end{split}
\end{equation}
which resembles the convention of \cite{Tagliavini2018,Rohringer2013a} when exploiting time translation invariance $\tau_4=0$, and further, the connected physical response function
\begin{equation}
\begin{split}
\chi^{2,\mathrm{b,f}}_{\text{ph},\sigma\sigma'}(\i\omega_m)&=\int_0^\beta \mathrm{d}\tau_1 \mathrm{d}\tau_4 \mathrm{e}^{\i\omega_m(\tau_4-\tau_1)}\\&\qquad\cdot\left[\langle T_{\tau}n^f_{\sigma}(\tau_1)n^\mathrm{b}_{\sigma'}(0)\rangle \right]\\&\qquad-n^\mathrm{f}_{\sigma} n^\mathrm{b}_{\sigma'}\delta(\i\omega)\\
&\overset{!}{=}\frac{\delta^{f,e}}{\beta}\sum_{n}\chi^{3,\mathrm{b,f,e}}_{\text{ph},\sigma\sigma'}(\i\omega_m,\i\nu_n),
\end{split}
\end{equation}
for both of which we define the charge and spin channels as in \cite{Tagliavini2018,Rohringer2012}:
\begin{equation}
\begin{split}
  &\chi_{\mathrm{ch}}=\chi_{\text{ph},\uparrow\uparrow}+\chi_{\text{ph},\uparrow\downarrow}, \\&\chi_{\mathrm{sp}}=\chi_{\text{ph},\uparrow\uparrow}-\chi_{\text{ph},\uparrow\downarrow}.
  \end{split}
\end{equation} 
The bosonic propagator in the specific channel is:
\begin{equation}
    \begin{split}
        w^{i,j}_{\mathrm{ch}}(\i\omega)&=U\delta^{i,j}-U^2\chi^{2,i,j}_{\mathrm{ch}}(\i\omega)\\
        w^{i,j}_{\mathrm{sp}}(\i\omega)&=U\delta^{i,j}+U^2\chi^{2,i,j}_{\mathrm{sp}}(\i\omega).
    \end{split}
\end{equation}
Subtracting the legs from the connected Fermi-Bose response functions yields the Hedin vertex $\lambda^{i,j,k}_{\mathrm{sp/ch}}$ in the specific channel:
\begin{equation}
    \begin{split}
        \lambda^{k,i,j}_{\mathrm{sp/ch}}(\i\omega,\i\nu)&=-\sum_{m,n,o}\frac{U\chi^{3,o,m,n}_{\mathrm{sp/ch}}(\i\omega,\i\nu)}{G^{m,i}(\i\nu)G^{j,n}(\i\nu+\i\omega)w^{o,k}_{\mathrm{sp/ch}}(\i\omega)},\\
    \end{split}
\end{equation} where the division indicates matrix inversion with respect to the orbital indices. We also reported this formulation in \cite{Kraemer2024}.

\subsection{Error estimate for the self-energy from the Hedin equation}
\label{App:Error}
\begin{figure}[t!]
    \centering
    \includegraphics[width=0.85\linewidth]{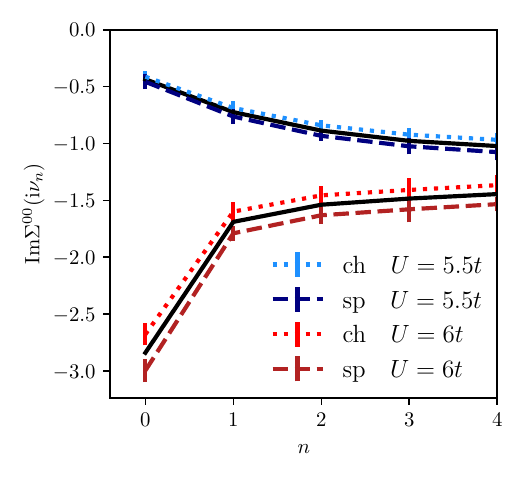}
    \caption{Error estimate for $U=5.5t$ (blue) and $U=6t$ (red) from the charge channel (dotted line) and spin channel (dashed line) in comparison to the directly measured self-energy (black lines). The error bars give the error of the self-energy from the statistic error of the Hedin vertices.}
    \label{fig:SigErr}
\end{figure}

Two major sources of errors can be determined: first, the finite grid of the Matsubara frequencies for the Hedin-vertex results underestimate correlations. Here, we used 82 positive fermionic frequencies and 81 positive bosonic frequencies.
 Fig.~\ref{fig:SigErr} displays the resulting self-energy for the charge- and spin channel for the renormalized metallic regime $U=5.5t$ (blue) and the insulating regime $U=6t$ (red) in comparison to the directly measured self-energy (black bold line). We observe for both cases, that the charge channel (dotted line) underestimates the self-energy, while the spin channel (dashed line) overestimates the self-energy, but to a qualitatively similar magnitude. This is reasoned by the high frequency tail of the Hedin vertex: While the tail of the Hedin vertices is similar for charge and spin Hedin vertex, they enter the self-energy of Eq.~(\ref{eq:Hedin}) with opposite sign.

Secondly, error-bars are given for the Hedin self-energy, which take into account statistical errors from the Hedin vertices by error propagation towards the self-energy. The Hedin vertices error were computed by the variance of multiple Hedin vertex measurements from different random seeds for the Monte Carlo processes, which resulted in the case of $U=6t$ in a relative error of $\Delta\lambda^{\mathrm{000}}_{\mathrm{sp}}/|\lambda^{\mathrm{000}}_{\mathrm{sp}}|\approx 0.02$ for the static bosonic frequency.

\subsection{Details of the CDMFT algorithm}
\label{App.CDMFT-Algorithm}
The CDMFT is a real space cluster extension~\cite{Parcollet2004,Maier2005} of the dynamical mean-field theory (DMFT,~\cite{Metzner1989,Georges1992,Georges1996}). In CDMFT the auxiliary impurity Anderson model does not correspond to a single lattice site but 
to a real space super-cell containing $N_c$ lattice sites, which converges to the lattice problem in the limit of infinitely large impurity clusters. For this study, we set $N_c=2\times2$, resulting in the excessively studied plaquette CDMFT, a sketch of which can be found in Fig.~\ref{fig:Hedin_example}. Single-particle level calculations for larger clusters can be found, e.g., in \cite{Klett2020,Downey2023,Meixner2024}.

Similarly to single-site DMFT, the auxiliary problem is determined by converging a self-consistency loop iteratively, so that the following condition is fulfilled:
\begin{equation}
\underbrace{\sum_{\Vec{k}\in\mathrm{RBZ}}\left[(\mathrm{i}\nu_n-\mu)\delta^{i,j}-\varepsilon^{i,j}(\Vec{k})-\Sigma^{i,j}(\mathrm{i}\nu_n)\right]^{-1}}_{G_{ext}^{i,j}}\,\overset{!}{=}G^{i,j}(\mathrm{i}\nu_n),\label{eq:self-consistency}
\end{equation} where $i,j$ are cluster-site indices, $\Vec{k}$ is the wave-vector in the reduced Brillouin zone (RBZ) of the super-lattice, $\varepsilon^{i,j}(\Vec{k})$ the lattice dispersion relation in momentum space and $\,^{-1}$ refers to a matrix-inversion with respect to the site-indices $i,j$, for fixed fermionic frequency $\i\nu$. In practice, the self-consistency condition Eq.~\eqref{eq:self-consistency} is achieved by starting with an initial guess of the self-energy and computing from this the impurity Weiss-field $\mathcal{G}_0(\mathrm{i}\nu_n)$ via a matrix-valued Dyson equation:
\begin{equation}
\left[\mathcal{G}_0^{-1}(\mathrm{i}\nu_n)\right]^{i,j}=\left[G_{\mathrm{ext}}^{-1}(\mathrm{i}\nu_n)\right]^{i,j}+\Sigma^{i,j}(\mathrm{i}\nu_n).
\end{equation} The Weiss-field then describes the influence of the fermionic dynamics of the surrounding bath to the Anderson impurity. Solving this impurity model results in an updated self-energy $\Sigma^{i,j}(\mathrm{i}\nu_n)$.

\begin{figure}[t!]
    \centering
    \includegraphics[width=\linewidth]{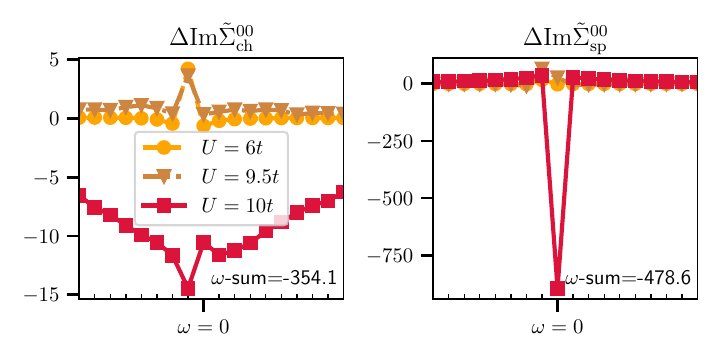}
    \caption{Spin- and charge fluctuation diagnostics for the self-energy difference of DMFT across the MIT from the charge- and spin channel and different interaction values. Note the different scales.}
    \label{fig:fd_DMFT}
\end{figure}
\section{Fluctuation diagnostics for the DMFT MIT}
\label{App:DMFT}

The local nature of single-site DMFT implicates that the fully local diagram in charge and spin channel reproduce the same (impurity) self-energy from the local diagrams. Fig.~\ref{fig:fd_DMFT} displays the fluctuation analysis of the difference of the self-energy in the charge channel in the first column and the spin channel in the second column at $T=0.067t$. For the renormalized metal up until $U=9.5t$, the charge and spin channel display a strong positive, metallic contribution from the minus-first bosonic frequency. Short after the MIT, at $U=10t$ (red full squares), the charge channel displays a broad frequency structure which contributes negatively and hence, insulating to the self-energy. This is different w.r.t. the spin channel, where the static fluctuations are strongly peaked insulating contributions, while all other fluctuations contribute slightly towards the metallic. The large difference in the $\omega$-sum between charge and spin channel results from a very slowly decaying high frequency tail in the charge channel, well beyond the frequency grid.
\begin{figure}[t!]
    \centering
    \includegraphics[width=\linewidth]{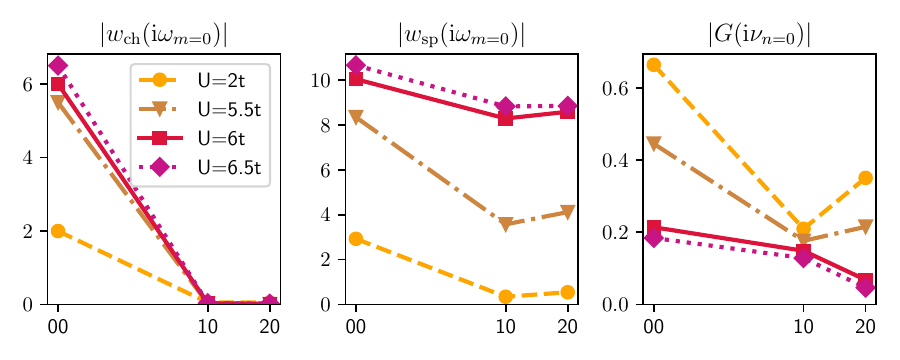}
    \caption{Decay of the absolute value of the charge- spin and fermionic propagator for different interaction values at $T=0.067t$. The possible distances covered by the propagator are sorted by their Euclidian distance.}
    \label{fig:Decay}
\end{figure}

\section{Further Results}
\subsection{Decay of the bosonic and fermionic propagator throughout the cluster}
\label{App:decay}
\begin{figure*}[t!]
    \centering
    \includegraphics[width=1\linewidth]{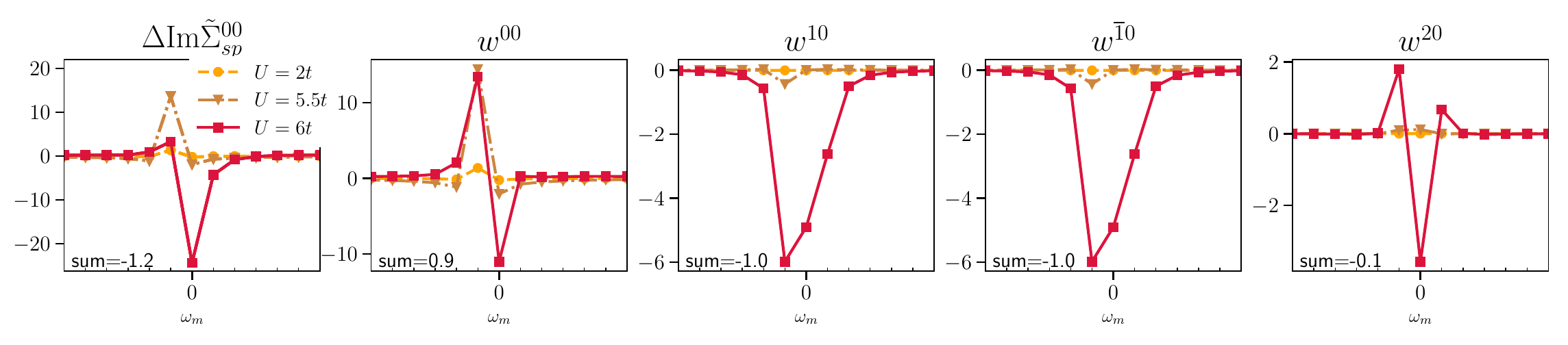}
    \caption{Real space fluctuation diagnostics of the real part of the nearest-neighbor self-energy's difference: The left panel shows the sum over all orbitals for different interactions. The four right panels show the respective real space contributions to the Hedin equation for a fixed bosonic propagator $w$ for the different interaction values. ``sum" gives the overall sum over all data points for $U=6t$ for the respective case.}
    \label{fig:FD_sigma_local_Udep}
\end{figure*}
Fig.~\ref{fig:Decay} discusses the decay of the absolute values of the static bosonic and fermionic propagators throughout the cluster and for different interaction values. Note that the local term of the bosonic propagator contains a bare interaction $U$. As becomes evident, the charge propagator is already purely local for the intermediate interaction regime, suppressing all non-local diagrams in the Hedin equation. This contrasts the bosonic spin propagator, which does barely decay throughout the cluster. Also, the fermionic propagator does not decay to the nearest-neighbor but is strongly suppressed to the second-nearest neighbor.

\subsection{Detailed interaction evolution of the on-site self-energy}

Fig.~\ref{fig:FD_sigma_local_Udep} gives the evolution of the difference of the self-energy for data before and after the MIT for diagram classes distinguish the locality of the bosonic propagator in the Hedin equation. While for the non-local contributions $w^{10},w^{\overline{1}0},w^{20}$, we find hardly a contribution before the MIT at $U=5.5t$, and strong insulating contributions in the cases of $U=6t$ and $6.5t$. For the nearest-neighboring components, we note, that at $U=6t$, close to the MIT, the static part is not dominating, as discussed in Sec.~\ref{sec:sig_loc}. This is overcome for $U=6.5t$. Taking a look now at the local contributions $w^{00}$, we note, that in the metallic case of $U=5.5t$, the high frequency components for $U=5.5t$ are negative, contributing towards the insulating, while the metal is actually protected by a large positive contribution for the $\i\omega_{n=-1}$-bosonic frequency. When crossing the MIT into the insulating regime at $U=6t$, the contribution of the higher frequency points turns rather counter-intuitively from negative, i.e. insulating to metallic.

 \subsection{Comparison to the single-boson irreducible self-energy decomposition}
 \label{App:SBE}
Our approach exploits the fact, that the Hubbard interaction $U$ can be arbitrarily split into a charge and spin interaction, which is the so-called Fierz-ambiguity \cite{Ayral2017,Krien2020c}, see App.~\ref{App:Derivation1}. This way, the fluctuation diagrams are reordered either in the charge- or spin picture but both channels contain all scattering processes and result in the same self-energy. This contrasts the recently proposed approach \cite{Yu2024} to exploit the single-boson irreducibility of the Hedin vertex: 
The self-energy is divided via the Hedin equation into single-boson exchange (SBE) diagrams in the charge, spin and particle-particle channel and a rest containing contributions from (up to second-order) perturbation theory and multi-boson exchange diagrams. We want to stress here, that the Hedin-like equation for the single-spin-boson diagram differs in the following ways from our spin-boson diagram:

(i) The diagram has a prefactor of 3/2 compared to our Hedin equation, and

(ii) The bosonic propagator in the SBE formalism does not contain the bare interaction, which is completely local.

\begin{figure*}[t!]
    \centering
    \includegraphics[width=1\linewidth]{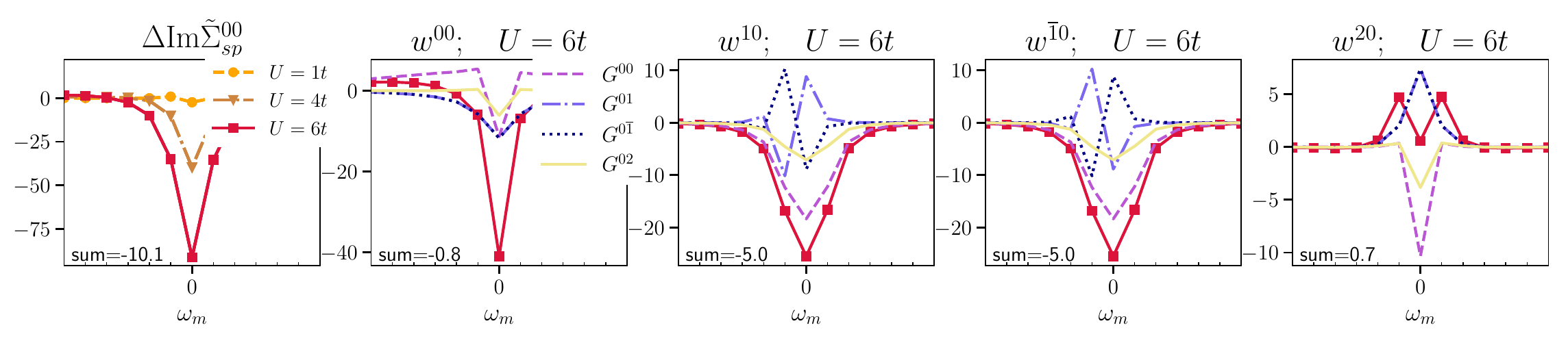}
    \caption{RFD in analogy to Fig.~\ref{fig:FD_sigma_local}, but for the imaginary part of the second-next neighbor self-energy of the Hubbard atom cluster.}
    \label{fig:FD_sigma_local_Udep_HUBBARD}
\end{figure*}

From this observation, and indeed respective calculations, we conclude the following: Our finding of an insulating, non-local contribution in the spin channel remains true for the single-boson exchange diagram - to be precise, the fluctuation-diagnostics from the functional $\omega$-dependence of $\tilde{\Sigma}^{00}\left[\lambda^{100}\right](\omega,\nu)$, are the same for the non-local single-boson exchange diagrams, only altered by a prefactor of 3/2. This prefactor results in a larger self-energy contribution in the SBE, which is in our case already cancelled by multi-boson diagrams in the spin channel.

\section{Real space fluctuation diagnostics for the local self-energy of the Hubbard atom \texorpdfstring{$2\times2$-}{}cluster}
\label{App:Hubbard_cluster}
As discussed in Fig.~\ref{fig:Hedin-to-sigma_local}, the Hedin vertex leading to the insulating contribution of the nearest-neighboring self-energy is of very similar appearance as the corresponding Hubbard atom clusters Hedin vertex. This leads directly to the question of fluctuation diagnostics of the Hubbard atom cluster, see Fig.~\ref{fig:FD_sigma_local_Udep_HUBBARD}. While we find a strong insulating contribution of the bosonic nearest-neighbor, fermionic local diagram $w_{10}$ and $w_{\overline{1}0}$, we further find an insulating, peaked structure for the static frequency of the fully local diagrams $w^{00}$, resulting for the Hubbard atom cluster in overall local insulting diagram.

\bibliography{main}
\end{document}